\documentclass{aa}
\usepackage{graphicx}
\usepackage{natbib}
\usepackage{bm}
\graphicspath{{./fig/}{./png/}}
\newcommand{\EQ}{\begin{equation}}
\newcommand{\EN}{\end{equation}}
\newcommand{\EQA}{\begin{eqnarray}}
\newcommand{\ENA}{\end{eqnarray}}
\newcommand{\eq}[1]{(\ref{#1})}

\newcommand{\Eq}[1]{Equation~(\ref{#1})}

\newcommand{\Fig}[1]{Figure~\ref{#1}}
\newcommand{\FFig}[1]{Figure~\ref{#1}}

\newcommand{\Figss}[2]{Figures~\ref{#1}--\ref{#2}}
\newcommand{\Tab}[1]{Table~\ref{#1}}

{}
{}
\newcommand{\meanFFFF}{\overline{\mbox{\boldmath ${\cal F}$}}{}}{}
\newcommand{\meanemf}{\overline{\cal E} {}}

{}
{}
\newcommand{\meanEMF}{\overline{\mbox{\boldmath ${\cal E}$}}{}}{}
{}
{}
{}
{}
{}
{}
\newcommand{\meanBB}{\overline{\mbox{\boldmath $B$}}{}}{}
{}
{}
{}
{}
{}
\newcommand{\meanGG}{\overline{\mbox{\boldmath $G$}}{}}{}
{}
{}
\newcommand{\meanJJ}{\overline{\mbox{\boldmath $J$}}{}}{}
\newcommand{\meanKK}{\overline{\mbox{\boldmath $K$}}{}}{}
\newcommand{\meanUU}{\overline{\bm{U}}}

{}
{}
{}

\newcommand{\meanB}{\overline{B}}

\newcommand{\meanC}{\overline{C}}

\newcommand{\meanJ}{\overline{J}}

\newcommand{\meanFFF}{\overline{\cal F}}

{}

{}
{}

\newcommand{\eee}{\hat{\mbox{\boldmath $e$}} {}}

\newcommand{\zzz}{\hat{\mbox{\boldmath $z$}} {}}

\newcommand{\nullvector}{{\bf0}}

\newcommand{\kk}{\bm{k}}

\newcommand{\xx}{\bm{x}}

\newcommand{\uu}{\mbox{\boldmath $u$} {}}
\newcommand{\UU}{\mbox{\boldmath $U$} {}}

\def\bb{\bm{b}}

\newcommand{\BB}{\mbox{\boldmath $B$} {}}

\newcommand{\JJ}{\mbox{\boldmath $J$} {}}

\newcommand{\aaaa}{\mbox{\boldmath $a$} {}}

\newcommand{\ff}{\mbox{\boldmath $f$} {}}

\newcommand{\grav}{\mbox{\boldmath $g$} {}}
\newcommand{\nab}{\mbox{\boldmath $\nabla$} {}}
\newcommand{\OO}{\bm{\Omega}}
\newcommand{\oo}{\mbox{\boldmath $\omega$} {}}

\newcommand{\xxi}{\mbox{\boldmath $\xi$} {}}
\newcommand{\SSSS}{\mbox{\boldmath ${\sf S}$} {}}

\newcommand{\ii}{{\rm i}}

\newcommand{\DD}{{\rm D} {}}

\newcommand{\dd}{{\rm d} {}}

\newcommand{\const}{{\rm const}  {}}

\newcommand{\sx}{\,{\rm s}x}
\newcommand{\sy}{\,{\rm s}y}
\newcommand{\sz}{\,{\rm s}z}
\newcommand{\cx}{\,{\rm c}x}
\newcommand{\cy}{\,{\rm c}y}
\newcommand{\cz}{\,{\rm c}z}
\def\Ma{\mbox{\rm Ma}}
\def\Co{\mbox{\rm Co}}

\def\Sc{\mbox{\rm Sc}}

\def\Gr{\mbox{\rm Gr}}
\def\Pm{P_{\rm m}}
\def\Rm{R_{\rm m}}

\def\Pe{\mbox{\rm Pe}}
\def\Co{\mbox{\rm Co}}

\def\cs{c_{\rm s}}

\def\kf{k_{\rm f}}

\def\urms{u_{\rm rms}}

\def\etat{\eta_{\rm t}}

\def\half{{\textstyle{1\over2}}}

\def\onethird{{\textstyle{1\over3}}}

\newcommand{\yapj}[3]{ #1, {ApJ,} {#2}, #3}

\newcommand{\yan}[3]{ #1, {Astron.\ Nachr.,} {#2}, #3}

\newcommand{\yana}[3]{ #1, {A\&A,} {#2}, #3}

\newcommand{\ygafd}[3]{ #1, {Geophys.\ Astrophys.\ Fluid Dyn.,} {#2}, #3}

\newcommand{\yprl}[3]{ #1, {Phys.\ Rev.\ Lett.,} {#2}, #3}

\newcommand{\ymn}[3]{ #1, {MNRAS,} {#2}, #3}

\newcommand{\ypr}[3]{ #1, {Phys.\ Rev.,} {#2}, #3}
\newcommand{\ypre}[3]{ #1, {Phys.\ Rev.\ E,} {#2}, #3}

\newcommand{\yjour}[4]{ #1, {#2}, {#3}, #4}

\newcommand{\sapj}[1]{ #1, {ApJ}, submitted}

\titlerunning{Mean-field transport in turbulence}

\title{Mean-field transport in stratified and/or rotating turbulence}
\author{A. Brandenburg\inst{1,2} \and K.-H. R\"adler\inst{3} \and K. Kemel\inst{1,2}}
\institute{
NORDITA, AlbaNova University Center,
Roslagstullsbacken 23, SE-10691 Stockholm, Sweden
\and
Department of Astronomy, AlbaNova University Center,
Stockholm University, SE-10691 Stockholm, Sweden
\and
Astrophysical Institute Potsdam, An der Sternwarte 16, D-14482 Potsdam, Germany
}
\date{\today,~ $ $Revision: 1.168 $ $}
\begin{document}

\abstract{
The large-scale magnetic fields of stars and galaxies are often described
in the framework of mean-field dynamo theory.
At moderate magnetic Reynolds numbers, the
transport coefficients defining the mean electromotive force
can be determined from simulations.
This applies analogously also to passive scalar transport.
}{
We investigate the mean electromotive force in the kinematic framework, that is,
ignoring the back-reaction of the magnetic field on the fluid velocity,
under the assumption of axisymmetric turbulence determined by the presence
of either rotation, density stratification, or both.
We use an analogous approach for the mean passive scalar flux.
As an alternative to convection, we consider forced turbulence in an isothermal layer.
When using standard ansatzes, the mean magnetic transport is then determined by nine,
and the mean passive scalar transport by four coefficients.
We give results for all these transport coefficients.
}{
We use the test-field method and the test-scalar method,
where transport coefficients are determined by solving sets
of equations with properly chosen mean magnetic fields or
mean scalars.
These methods are adapted to mean fields
which may depend on all three space coordinates.
}{
We find the anisotropy of turbulent diffusion to be moderate
in spite of rapid rotation or strong density stratification.
Contributions to the mean electromotive force determined by
the symmetric part of the gradient tensor of the
mean magnetic field, which were ignored in several earlier investigations,
turn out to be important.
In stratified rotating turbulence, the $\alpha$ effect is strongly
anisotropic, suppressed along the rotation axis on large length scales,
but strongly enhanced at intermediate length scales.
Also the $\OO\times\meanJJ$ effect is enhanced at intermediate length scales.
The turbulent passive scalar diffusivity is typically almost twice as
large as the turbulent magnetic diffusivity.
Both magnetic and passive scalar diffusion are slightly enhanced
along the rotation axis, but decreased if there is gravity.
}{
The test-field and test-scalar methods provide powerful tools
for analyzing transport properties of axisymmetric turbulence.
Future applications are proposed ranging from anisotropic turbulence due
to the presence of a uniform magnetic field to inhomogeneous turbulence
where the specific entropy is nonuniform, for example.
Some of the contributions to the mean electromotive force which have been ignored
in several earlier investigations, in particular those given by the symmetric part
of the gradient tensor of the mean magnetic field, turn out to be of significant magnitude.
}

\maketitle

\section{Introduction}

Stellar mixing length theory is a rudimentary description of turbulent convective energy transport.
The mixing length theory of turbulent transport goes back to \cite{Pra25}
and, in the stellar context, to \cite{Vit53}.
The simplest form of turbulent transport is turbulent diffusion, which
quantifies the mean flux of a given quantity,
e.g., momentum, concentration of chemicals, specific entropy or magnetic fields,
down the gradient of its mean value.
In all these cases essentially a Fickian diffusion law is established,
where the turbulent diffusion coefficient is proportional to the rms velocity of the turbulent eddies
and the effective mean free path of the eddies or their correlation length.

Mean-field theories, which have been elaborated, e.g., for the behavior of magnetic fields
or of passive scalars in turbulent media, go beyond this concept.
In the case of magnetic fields, the effects of turbulence occur in a mean electromotive force,
which is related to the mean magnetic field and its derivatives in a tensorial fashion.
Examples for effects described by the mean magnetic field alone,
without spatial derivatives, are the $\alpha$-effect \citep{SKR66}
and the pumping of mean magnetic flux \citep{Rae66,Rae68,RS75};
for more information on these topics see, e.g., \cite{KR80} or \cite{BS05}.
Likewise the mean passive scalar flux contains a pumping effect \citep{EKR96}.
In both the magnetic and the passive scalar cases turbulent diffusion occurs, which is in general anisotropic.
The coupling between the mean electromotive force and the magnetic field and its derivatives,
or mean passive scalar flux and the mean scalar and its derivatives, is given by turbulent transport coefficients.

On the analytic level of the theory the determination of these transport coefficients is only possible
with some approximations.
The most often used one is the second-order correlation approximation (SOCA),
which has delivered so far many important results.
Its applicability is however restricted to certain ranges of parameters
like the magnetic Reynolds number or the P\'eclet number.
In spite of this restriction, SOCA is an invaluable tool,
because it allows a rigorous treatment
within the limits of its applicability.
It is in particular important for testing numerical methods that apply in a wider range.

In recent years it has become possible to compute the full set of turbulent
transport coefficients numerically from simulations of turbulent flows.
The most accurate method for that is the test-field method \citep{Sch05,Sch07}.
In addition to the equations describing laminar and turbulent flows,
one solves a set of evolution equations for the small-scale magnetic
or scalar fields which result from given mean fields, the test fields.
By selecting a sufficient number of independent test fields, one obtains
a corresponding number of mean electromotive forces or mean scalar fluxes
and can then compute in a unique way
all the associated transport coefficients.

Most of the applications of the test-field method are based
on spatial averages that are taken over two coordinates.
In the magnetic case this
approach has been applied to a range of different flows
including isotropic homogeneous turbulence \citep{SBS08,BRS08},
homogeneous shear flow turbulence \citep{BRRK08}
without and with helicity \citep{Mitra09},
and turbulent convection \citep{KKB09}.
One of the main results is that in the isotropic case, for
magnetic Reynolds numbers $\Rm$ larger than unity, the turbulent
diffusivity is given by $\onethird\tau\urms^2$, where the correlation
time $\tau$ is, to a good approximation, given by $\tau=(\urms\kf)^{-1}$.
Here, $\urms$ is the rms velocity of the turbulent small-scale flow
and $\kf$ is the wavenumber of the energy-carrying eddies.
For smaller $\Rm$, the turbulent diffusivity grows linearly with $\Rm$.
Furthermore, if the turbulence is driven isotropically by polarized waves,
the flow becomes helical and there is an $\alpha$ effect.
In the kinematic regime (for weak magnetic fields),
the $\alpha$ coefficient is proportional to
$\overline{\oo\cdot\uu}$, where $\oo=\nab\times\uu$
is the vorticity of the small-scale flow, $\uu$.
In the passive scalar case,
test scalars are used to determine the transport coefficients.
Results have been obtained for anisotropic flows in the presence of
rotation or strong magnetic fields \citep{BSV09}, linear shear \citep{MB10},
and for irrotational flows \citep{RBSR11}.

The present paper deals with the magnetic and the passive scalar case in the above sense.
Its goal is to compute the transport coefficients for axisymmetric turbulence,
that is, turbulence with one preferred direction, given by the presence of
either rotation or density stratification
or, if the relevant directions coincide, of both.
(Axisymmetric turbulence can be defined by requiring that any averaged
quantity depending on the turbulent velocity field is invariant under
any rotation of this field about the preferred axis.)
Note that a dynamo-generated magnetic field will in general violate the
assumption of axisymmetric turbulence.
To avoid this problem while still being able to investigate the general
effects arising from only one preferred direction, we assume
such fields to be weak so as not to affect the assumption of
axisymmetry of the turbulence.
An imposed uniform magnetic field in the preferred direction
would still be allowed, but this case will not be investigated in this paper;
see \cite{BSV09} for numerical investigations of passive scalar transport
with a uniform field.

Except for a few comparison cases, we always consider flows in a slab
between stress-free boundaries.
This is the simplest example of flows that are non-vanishing on the boundary and
compatible with axisymmetric turbulence.
To facilitate comparison with earlier work on forced turbulence,
we consider an isothermal layer even in the density-stratified case,
i.e., there is no convection,
and the flow is driven by a prescribed random forcing.
This is similar to earlier work on forced homogeneous turbulence
\citep{BRS08,BRRK08,BSV09}, but now
we will be able to address questions regarding vertical pumping
as well as helicity production and $\alpha$ effect in the presence of
rotation.
This setup allows us to isolate effects of density stratification from
those originating from the nonuniformities of turbulence intensity and
local correlation length.
In addition to isothermal stratification, we assume an isothermal equation
of state and thus do not consider an equation for the specific entropy.
Hence, no Brunt-V\"ais\"al\"a oscillations can occur.
This assumption would need to be relaxed for studying turbulent convection,
which will be the subject of a future investigation.

\section{Mean-field concept in turbulent transport}

\subsection{Mean electromotive force}

The evolution of the magnetic field $\BB$ in an electrically
conducting fluid is assumed to obey the induction equation,
\EQ
{\partial\BB\over\partial t}=\nab\times\left(
\UU\times\BB-\eta\JJ\right),
\label{eq001}
\EN
where $\UU$ is the velocity and $\eta$ the microscopic magnetic diffusivity
of the fluid, and $\JJ$ is defined by $\JJ = \nab\times\BB$
(so that $\JJ / \mu_0$ with $\mu_0$ being the magnetic permeability
is the electric current density).
We define mean fields as averages, assume that the averaging satisfies
(exactly or approximately) the Reynolds rules,
and denote averaged quantities by overbars.\footnote{The Reynolds rules
imply that $\overline{F + G} = \overline{F} + \overline{G}$,
$\overline{\overline{F}} = \overline{F}$, $\overline{\overline{F} G} = \overline{F} \overline{G}$,
$\overline{\partial F / \partial x} = \partial \overline{F} / \partial x$
and $\overline{\partial F / \partial t} = \partial \overline{F} / \partial t$
for any fluctuating quantities $F$ and $G$.}
The mean magnetic field $\meanBB$ is then governed by
\EQ
{\partial\meanBB\over\partial t}=\nab\times\left(
\meanUU\times\meanBB+\meanEMF-\eta\meanJJ\right),
\label{eq003}
\EN
where $\meanEMF=\overline{\uu\times\bb}$ is the mean electromotive
force resulting from the correlation of velocity and magnetic field
fluctuations, $\uu=\UU-\meanUU$ and $\bb=\BB-\meanBB$.

We focus attention on the mean electromotive force $\meanEMF$ in cases in which the velocity fluctuations $\uu$
constitute axisymmetric turbulence, that is, turbulence with one preferred direction,
which we describe by the unit vector $\eee$.
Until further notice
we accept the traditional assumption according to which $\meanEMF$ in a given point in space and time
is a linear homogeneous function of $\meanBB$ and its first spatial derivatives in this point.
Then, $\meanEMF$ can be represented in the form
\EQA
\meanEMF&=&
-\alpha_\perp\meanBB
-(\alpha_\parallel-\alpha_\perp)(\eee\cdot\meanBB)\eee
-\gamma\eee\times\meanBB
\nonumber \\
&&-\beta_\perp\meanJJ
-(\beta_\parallel-\beta_\perp)(\eee\cdot\meanJJ)\eee
-\delta\eee\times\meanJJ
\label{eq005}\\
&&-\kappa_\perp\meanKK
-(\kappa_\parallel-\kappa_\perp)(\eee\cdot\meanKK)\eee
-\mu\eee\times\meanKK
\nonumber
\ENA
with nine coefficients $\alpha_\perp$, $\alpha_\parallel$, $\ldots$, $\mu$.\footnote{Note
that the signs in front of some individual terms on the right-hand side of \eq{eq005},
in particular of those with $\alpha_\perp$ and $\alpha_\parallel$
(perpendicular and parallel $\alpha$ effect) as well as $\gamma$
(pumping in the $z$ direction),
may differ from the signs used in other representations.}
Like $\meanJJ = \nab \times \meanBB$, also $\meanKK$ is determined by the gradient tensor $\nab \meanBB$.
While $\meanJJ$ is given by its antisymmetric part,
$\meanKK$ is a vector defined by $\meanKK=\eee \cdot (\nab\meanBB)^\mathrm{S}$
with $(\nab\meanBB)^\mathrm{S}$ being the symmetric part of $\nab \meanBB$.
A more detailed explanation of \eq{eq005} is given in Appendix \ref{app1}.
If $\eee$ is understood as polar vector
(for example $\nab \overline{\rho} / |\nab \overline{\rho}|$,
where $\overline{\rho}$ is the mean mass density),
then $\meanKK$ is axial and $\gamma$, $\beta_\perp$, $\beta_\parallel$
and $\mu$ are true scalars, but $\alpha_\perp$, $\alpha_\parallel$,
$\delta$, $\kappa_\perp$ and $\kappa_\parallel$ pseudoscalars.
(Scalars are invariant but pseudoscalars change sign if the turbulent velocity field is reflected
at a point or at a plane containing the preferred axis.)
Sometimes it is useful to interpret $\eee$ as an axial vector
(for example $\OO / |\OO|$ with $\OO$ being an angular velocity).
Then, $\meanKK$ is a polar vector, $\beta_\perp$, $\beta_\parallel$,
$\delta$, $\kappa_\perp$, $\kappa_\parallel$
and $\mu$ are true scalars but $\alpha_\perp$, $\alpha_\parallel$ and $\gamma$ pseudoscalars.

We may split $\meanEMF$ and $\meanBB$ into parts $\meanEMF_\perp$ and $\meanBB_\perp$ perpendicular to $\eee$
and parts $\meanEMF_\parallel$ and $\meanBB_\parallel$ parallel to it.
Then \eq{eq005} can be written in the form
\EQA
\meanEMF_\perp &=&
-\alpha_\perp \meanBB_\perp
-\gamma\eee \times \meanBB_\perp
-\beta_\perp \meanJJ_\perp
-\delta \eee \times \meanJJ_\perp
\nonumber\\
&&-\kappa_\perp \meanKK_\perp
-\mu \eee \times \meanKK_\perp
\label{eq007}\\
\meanEMF_\parallel &=&
- \alpha_\parallel \meanBB_\parallel
-\beta_\parallel \meanJJ_\parallel
-\kappa_\parallel \meanKK_\parallel \, .
\nonumber
\ENA

Let us return to \eq{eq005}.
In the simple case of homogeneous isotropic turbulence we have $\alpha_\perp = \alpha_\parallel$
and $\beta_\perp = \beta_\parallel$, and all remaining coefficients vanish.
Then, \eq{eq005} takes the form $\meanEMF = \alpha \meanBB - \etat \meanJJ$
with properly defined $\alpha$ and $\etat$.
These two coefficients have been determined by test-field calculations
\citep{SBS08,BRS08}.

In several previous studies of $\meanEMF$, more general kinds of turbulence
(that is, not only axisymmetric turbulence) have been considered,
but with a less general definition of mean fields, which were just
horizontal averages.
More precisely, Cartesian coordinates $(x, y, z)$ were adopted
and the averages were taken over all $x$ and $y$ so that they
depend on $z$ and $t$ only \citep{BRS08,BRRK08}.
This definition implies remarkable simplifications.
Of course, we then have $\meanJ_z = 0$.
Further, there are no non-zero components of $\nab \meanBB$ other
than $\meanB_{x,z}$ and $\meanB_{y,z}$, for $\nab \cdot \meanBB = 0$ requires $\meanB_{z,z} = 0$,
and these components can be expressed as components of $\meanJJ$,
{\it viz.} $\meanB_{x,z} = \meanJ_y$ and $\meanB_{y,z} = - \meanJ_x$.
(Here and in what follows, commas denote partial derivatives.)
This again implies $\meanKK = - {\textstyle{1 \over 2}} \, \eee \times \meanJJ$.
As a consequence, this definition of mean fields reduces (\ref{eq005}) to
\EQA
\meanEMF &=& - \alpha_\perp \meanBB - (\alpha_\parallel-\alpha_\perp)(\eee \cdot \meanBB) \eee
    - \gamma \eee \times \meanBB
\nonumber\\
&& -\beta^\dag \meanJJ - \delta^\dag \, \eee \times \meanJJ \, ,
\label{eq011}
\ENA
where $\beta^\dag = \beta_\perp + {\textstyle{1 \over 2}} \mu$
and $\delta^\dag = \delta - {\textstyle{1 \over 2}} \kappa_\perp$.
Of course, $\alpha_\perp$, $\alpha_\parallel$, $\gamma$, $\beta^\dag$ and $\delta^\dag$
are independent of $x$ or $y$.
Clearly, $\beta_\perp$ and $\mu$ as well as $\delta$ and $\kappa_\perp$
have no longer independent meanings.
From \eq{eq003} we may conclude that $\partial \meanB_z / \partial t = 0$.
If we restrict ourselves to applications in which $\meanB_z$ vanishes
initially, it does so at all times
and the term with $\alpha_\parallel-\alpha_\perp$ in \eq{eq011} disappears.
Then, only the four coefficients $\alpha_\perp$, $\gamma$, $\beta^\dag$
and $\delta^\dag$ are of interest.
They can be determined by test-field calculations using two test fields
independent of $x$ and $y$ \citep{BRS08,BRRK08}.

In this paper we go beyond the aforementioned assumptions
in the following respects.
Firstly, we relax the assumption that $\meanEMF$ in a given point in space is a homogeneous function of $\meanBB$
and its first spatial derivatives in this point.
Instead, we admit a non-local connection between $\meanEMF$ and $\meanBB$.
For simplicity, however, we further on
assume that $\meanEMF$ at a given time depends only on $\meanBB$ at the same time,
that is, we remain with an instantaneous connection between $\meanEMF$ and $\meanBB$.
This approximation requires that the mean field varies slowly on a time scale
much longer than the turnover time of the turbulence; see \cite{HB09} for a
more general treatment of rapidly changing fields.
Secondly, we consider mean fields no longer as averages over all $x$ and $y$.
We define $\meanBB$ at a point $(x, y)$ in a plane $z = \const$ by
averaging over some surroundings of this point in this plane so that it
still depends on $x$ and $y$.
In that sense we generalize \eq{eq005} so that
\EQA
\meanEMF (\xx) &=& - \int \!\! \big( \alpha_\perp (\xx,\xxi) \meanBB (\xx - \xxi)
\nonumber\\
&& \qquad \quad + \big(\alpha_\parallel (\xx,\xxi) - \alpha_\perp (\xx,\xxi) \big) \big(\eee \cdot \meanBB (\xx - \xxi) \big) \eee
\nonumber\\
&& \qquad \quad + \gamma (\xx,\xxi) \, \eee \times \meanBB (\xx - \xxi)
\nonumber\\
&& \qquad \quad + \beta_\perp (\xx,\xxi) \, \meanJJ (\xx - \xxi)
\nonumber\\
&& \qquad \quad + \big( \beta_\parallel (\xx,\xxi) - \beta_\perp (\xx,\xxi) \big) \big(\eee \cdot \meanJJ (\xx - \xxi) \big) \eee
\nonumber\\
&& \qquad \quad + \delta (\xx,\xxi) \, \eee \times \meanJJ (\xx - \xxi)
\label{eq013}\\
&& \qquad \quad + \kappa_\perp (\xx,\xxi) \, \meanKK (\xx - \xxi)
\nonumber\\
&& \qquad \quad + \big( \kappa_\parallel (\xx,\xxi) - \kappa_\perp (\xx,\xxi) \big) \big(\eee \cdot \meanKK (\xx - \xxi) \big) \eee
\nonumber\\
&& \qquad \quad + \mu (\xx,\xxi) \, \eee \times \meanKK (\xx - \xxi) \big) \, \dd^3 \xi \, .
\nonumber
\ENA
As a consequence of the axisymmetry of the turbulence, the coefficients
$\alpha_\perp$, $\alpha_\parallel$, $\ldots$, $\mu$
depend only via $\xi_x^2 + \xi_y^2$ on $\xi_x$ and $\xi_y$.
We consider them also as symmetric in $\xi_z$.
The integration is over all $\xxi$ space.
Of course, $\meanEMF$, $\meanBB$, $\meanJJ$, and $\meanKK$ may depend on $t$.
For simplicity, however, the argument $t$ has been dropped.

Let us subject \eq{eq013} to a Fourier transformation with respect to $\xxi$.
We define it by
\EQ
F (\xxi) = (2 \pi)^{-3} \int \tilde{F} (\kk) \, \exp( \ii \kk \cdot \xxi ) \, \dd^3 k \, .
\label{eq015}
\EN
Remembering the convolution theorem we obtain
\EQA
&&\meanEMF (\xx) = - (2\pi)^{-3} \int \Big(
\tilde{\alpha}_\perp (\xx,\kk) \tilde{\meanBB} (\kk)
\nonumber\\
&& \quad
    + \big(\tilde{\alpha}_\parallel (\xx,\kk) - \tilde{\alpha}_\perp (\xx,\kk) \big) \big(\eee \cdot \tilde{\meanBB} (\kk) \big) \, \eee
\nonumber\\
&& \quad + \tilde{\gamma} (\xx,\kk) \, \eee \times \tilde{\meanBB} (\kk)
\nonumber\\
&& \quad + \tilde{\beta}_\perp (\xx,\kk) \, \tilde{\meanJJ} (\kk)
    + \big(\tilde{\beta}_\parallel (\xx,\kk) - \tilde{\beta}_\perp (\xx,\kk) \big) \big(\eee \cdot \tilde{\meanJJ} (\kk) \big) \, \eee
\nonumber\\
&& \quad + \tilde{\delta} (\xx,\kk) \, \eee \times \tilde{\meanJJ} (\kk)
\label{eq017}\\
&& \quad + \tilde{\kappa}_\perp (\xx,\kk) \tilde{\meanKK} (\kk)
    + \big(\tilde{\kappa}_\parallel (\xx,\kk) - \tilde{\kappa}_\perp (\xx,\kk) \big) \big(\eee \cdot \tilde{\meanKK} (\kk) \big) \, \eee
\nonumber\\
&& \quad + \tilde{\mu} (\xx,\kk) \, \eee \times \tilde{\meanKK} (\kk) \Big) \, \exp (\ii \kk \cdot \xx) \, \dd^3\kk \, ;
\nonumber
\ENA
see \cite{CMRB11} for a corresponding relation in the case
of horizontally averaged magnetic fields that depend only on $z$.
Like $\alpha_\perp$, $\alpha_\parallel$, $\ldots$, $\mu$,
the $\tilde{\alpha}_\perp$, $\tilde{\alpha}_\parallel$, $\ldots$, $\tilde{\mu}$
are real quantities.
They depend only via $k_\perp = (k_x^2 + k_y^2)^{1/2}$ on $k_x$ and $k_y$
and are symmetric in $k_z$, i.e., depend only via $k_\parallel =|k_z|$ on $k_z$.
As $\alpha_\perp$, $\alpha_\parallel$, $\ldots$, $\mu$ are real
and symmetric in $\xi_x$, $\xi_y$ and $\xi_z$ we have
\EQ
\tilde{\alpha}_\perp (\xx, \kk)
     = \int \alpha_\perp (\xx, \xxi) \, \cos k_x \xi_x \, \cos k_y \xi_y \, \cos k_z \xi_z \, \dd^3 \xi
\label{eq019}
\EN
and analogous relations for $\tilde{\alpha}_\parallel$, $\ldots$, $\tilde{\mu}$.
We note that $\tilde{\alpha}_\perp$, $\ldots$, $\tilde{\mu}$, taken
at $\kk = \nullvector$, agree with $\alpha_\perp$, $\ldots$, $\mu$ in \Eq{eq005}.

\subsection{Mean passive scalar flux}

There are interesting analogies between turbulent transport of magnetic flux and that
of a passive scalar \citep[cf.][]{RBSR11}.
Assume that the evolution of a passive scalar $C$, e.g., the concentration of an admixture in a fluid,
is given by
\EQ
{\partial C\over\partial t}=-\nab\cdot(\UU C-D\nab C),
\label{eq021b}
\EN
where $D$ is the microscopic (molecular) diffusivity.
Then the mean scalar $\meanC$ has to satisfy
\EQ
{\partial\meanC\over\partial t}=-\nab\cdot(\meanUU \, \meanC+\meanFFFF-D\nab\meanC),
\label{eq023b}
\EN
where $\meanFFFF=\overline{\uu c}$ is the mean passive scalar flux,
$\uu$ stands again for the fluctuations of the velocity and  $c=C-\meanC$ for the fluctuations of $C$.
Consider again axisymmetric turbulence with a preferred direction given by the unit vector $\eee$.
Assume that $\meanFFFF$ in a given point in space and time is determined
by $\meanC$ and its gradient $\meanGG = \nab \meanC$ in this point.
Then we have
\EQ
\meanFFFF=-\gamma^C\meanC \eee
-\beta^C_\perp\meanGG
-(\beta^C_\parallel-\beta^C_\perp)(\eee\cdot\meanGG)\eee
-\delta^C\eee\times\meanGG,
\label{eq025}
\EN
with coefficients $\gamma^C$, $\beta^C_\perp$, $\beta_\parallel^C$ and $\delta^C$.
If $\eee$ is a polar vector, $\gamma^C$ is a scalar but $\delta^C$ a pseudoscalar,
and if $\eee$ is an axial vector, $\gamma^C$ is a pseudoscalar but $\delta^C$ a scalar,
while $\beta^C_\perp$ and $\beta_\parallel^C$ are always scalars.
We note that $\nab \cdot (\delta^C \eee \times \meanGG)$ is only unequal zero if $\delta^C$ is not constant
but varies in the direction of $\eee \times \meanGG$.

We may split $\meanFFFF$ and $\meanGG$ into parts
$\meanFFFF_\perp$ and $\meanGG_\perp$ perpendicular to $\eee$, and parts
$\meanFFFF_\parallel$ and $\meanGG_\parallel$ parallel to it, and
give \eq{eq025} the form
\EQA
\meanFFFF_\perp\!&=\!&
-\beta^C_\perp\meanGG_\perp-\delta^C\eee\times\meanGG_\perp
\nonumber\\
\meanFFFF_\parallel\!&=\!&
-\gamma^C\eee\meanC-\beta^C_\parallel\meanGG_\parallel.
\label{eq026}
\ENA

Let us now relax the assumption that $\meanFFFF$ in a given point in space and time
is determined by $\meanC$ and $\meanGG$ in this point.
Analogously to the magnetic case we consider a non-local but instantaneous connection
between $\meanFFFF$ and $\meanC$.
Then we have
\EQA
\meanFFFF (\xx) &=& - \int \Big( \gamma^C (\xx, \xxi) \, \eee \, \meanC (\xx - \xxi)
\nonumber\\
&& \qquad + \beta^C_\perp (\xx, \xxi) \meanGG (\xx - \xxi)
\nonumber\\
&& \qquad + \big( \beta^C_\parallel (\xx, \xxi) - \beta^C_\perp (\xx, \xxi) \big) \,
     \big( \eee \cdot \meanGG (\xx - \xxi) \big) \, \eee
\label{eq029}\\
&& \qquad + \delta^C (\xx, \xxi) \, \eee \times \meanGG (\xx - \xxi) \Big) \, \dd^3\xi \, .
\nonumber
\ENA
As $\alpha_\perp$, $\alpha_\parallel$, $\ldots$, $\mu$ in the magnetic case,
$\gamma^C$, $\beta^C_\perp$, $\beta_\parallel^C$ and $\delta^C$
depend only via $\xi_x^2 + \xi_y^2$ on $\xi_x$ and $\xi_y$,
and we consider them also as symmetric in $\xi_z$.
The integration is again over all $\xxi$ space.
Note that $\meanFFFF$, $\meanC$, and $\meanGG$ may,
even if it is not explicitly indicated, depend on $t$.
Applying the Fourier transformation defined by \eq{eq015} on \eq{eq029},
we arrive at
\EQA
\meanFFFF (\xx) &=& - (2\pi)^{-3} \int \Big( \tilde{\gamma}^C (\xx, \kk) \, \eee \, \tilde{\meanC} (\kk)
\nonumber\\
&& \qquad + \tilde{\beta}^C_\perp (\xx, \kk) \tilde{\meanGG} (\kk)
\nonumber\\
&& \qquad + \big( \tilde{\beta}^C_\parallel (\xx, \kk) - \tilde{\beta}^C_\perp (\xx, \kk) \big) \,
     \big( \eee \cdot \tilde{\meanGG} (\kk) \big) \, \eee
\label{eq031}\\
&& \qquad + \tilde{\delta}^C (\xx, \kk) \, \eee \times \tilde{\meanGG} (\kk) \Big) \,
     \exp( \ii \kk \cdot \xx) \, \dd^3k \, ,
\nonumber
\ENA
where $\tilde{\gamma}^C_\perp$, $\tilde{\beta}^C_\perp$, $\tilde{\beta}^C_\parallel$ and $\tilde{\delta}^C$
are real quantities.
They depend only via $k_x^2 + k_y^2$ on $k_x$ and $k_y$,
and only via $k_\parallel$ on $k_z$, and they
satisfy relations analogous to \eq{eq019}.
We note that $\tilde{\gamma}^C$, $\tilde{\beta}^C_\perp$,
$\tilde{\beta}^C_\parallel$, and $\tilde{\delta}^C$
at $\kk = \nullvector$ agree with $\gamma^C$, $\beta^C_\perp$,
$\beta^C_\parallel$, and $\delta^C$
in \eq{eq025}.

\section{Simulating the turbulence}

We assume that the fluid is compressible and its flow is governed by the equations
\EQA
{\DD\UU\over\DD t} &=& \ff+\grav-\nab h-2\OO\times\UU
+\rho^{-1}\nab\cdot(2\nu\rho\SSSS)
\nonumber\\
{\DD h\over\DD t} &=&-\cs^2\nab\cdot\UU \, .
\label{eq041}
\ENA
Here, $\ff$ means a random force which primarily
drives isotropic turbulence \citep[e.g.,][]{HBD04},
$\grav$ the gravitational force,
and $h$ the specific enthalpy.
An isothermal equation of state, $p=\rho\cs^2$, has been adopted
with a constant isothermal sound speed $\cs$.
In general a fluid flow in a rotating system is considered,
$\OO$ is the angular velocity which defines the Coriolis force.
As usual $\rho$ means the mass density, $\nu$ the kinematic viscosity
and $\SSSS$ the trace-free rate of strain tensor,
${\sf S}_{ij} = \half (U_{i,j} + U_{j,i})
-\onethird\delta_{ij}\nab\cdot\UU$.
The influence of the magnetic field on the fluid motion, that is the
Lorentz force, is ignored throughout the paper.

The numerical simulation is carried out in a cubic domain of size $L^3$,
so the smallest wavenumber is $k_1=2\pi/L$.
In most of the cases a density stratification is included with $\grav = (0, 0, -g)$,
so the density scale height is $H_\rho=\cs^2/g$.
The number of scale heights across the domain is equal to $\Delta\ln\rho$,
where $\Delta$ denotes the difference of values at the two edges of the domain.
The forcing is assumed to work with an average wavenumber $\kf$.
The scale separation ratio is then given by $\kf/k_1$,
for which we usually adopt the value 5.
This means that we have about 5 eddies in each of the three coordinate directions.

The flow inside the considered domain depends on the boundary conditions.
Unless indicated otherwise we take the top and bottom surfaces
$z=z_1$ and $z=z_2$ with $z_2=-z_1=L/2$ as stress-free
and adopt periodic boundary conditions for the other surfaces.

\section{Computing the transport coefficients}

\subsection{Test-field method}
\label{sec32}

In the magnetic case the coefficients $\alpha_\perp$, $\alpha_\parallel$, $\ldots$, $\mu$
are determined by the test-field method \citep{Sch05,Sch07,BRS08}.
This method works with a set of test fields $\meanBB$, called $\meanBB^{\rm{T}}$,
and the corresponding mean electromotive forces $\meanEMF$, called $\meanEMF^{\rm{T}}$.
For the latter we have $\meanEMF^{\rm{T}} = \overline{\uu \times \bb^{\rm{T}}}$,
where the $\bb^{\rm{T}}$ obey
\EQA
\bb^T &=& \nab\times\aaaa^T
\nonumber\\
{\partial\aaaa^T\over\partial t} &=& \meanUU\times\bb^T+\uu\times\meanBB^T+
(\uu\times\bb^T)'+\eta\nabla^2\aaaa^T \, ,
\label{eq051}
\ENA
with $\meanUU$ and $\uu$ taken from the solutions of \eq{eq041}.
For the boundaries $z=\const$ we choose conditions which correspond to an adjacent perfect conductor,
for the $x$ and $y$ directions periodic boundary conditions.

We define four test fields by
\EQA
\meanBB^{\rm 1s} &=& (B_0 \sx \sy \sz, 0, 0) \, , \quad \meanBB^{\rm 1c} = (B_0 \sx \sy \cz, 0, 0)
\nonumber\\
\meanBB^{\rm 2s} &=& (0, 0, B_0 \sx \sy \sz) \, , \quad \meanBB^{\rm 2c} = (0, 0, B_0 \sx \sy \cz)
\label{eq053}
\ENA
with a constant $B_0$.
Here and in what follows we use the abbreviations
\EQA
\sx &=& \sin k_x x \, , \quad  \cx = \cos k_x x
\nonumber\\
\sy &=& \sin k_y y \, , \quad \cy = \cos k_y y
\label{eq055}\\
\sz &=& \sin k_z z \, , \quad \cz = \cos k_z z \, .
\nonumber
\ENA
We recall that test-fields need not to be solenoidal \citep[see][]{Sch05,Sch07}.

We denote the mean electromotive forces which correspond to the test fields \eq{eq053}
by $\meanEMF^{\rm 1s}$, $\meanEMF^{\rm 1c}$, $\meanEMF^{\rm 2s}$, and $\meanEMF^{\rm 2c}$.
With the presentation \eq{eq013} and relations like \eq{eq019} we find
\EQA
\meanemf^{\rm 1s}_x &=& - B_0 \big(\tilde{\alpha}_\perp \sx \sy \sz
    - (\tilde{\delta} - \frac{1}{2} \tilde{\kappa}_\perp) k_z \sx \sy \cz \big)
\nonumber\\
\meanemf^{\rm 1s}_y &=& - B_0 \big(\tilde{\gamma} \sx \sy \sz
    + (\tilde{\beta}_\perp + \frac{1}{2} \tilde{\mu}) k_z \sx \sy \cz  \big)
\nonumber\\
\meanemf^{\rm 1s}_z &=&  B_0 \, \tilde{\beta}_\parallel k_y \sx \cy \sz
\label{eq057}\\
\meanemf^{\rm 2s}_x &=& - B_0 \big( (\tilde{\beta}_\perp - \frac{1}{2} \tilde{\mu}) k_y \sx \cy \sz
     + (\tilde{\delta} + \frac{1}{2} \tilde{\kappa}_\perp) k_x \cx \sy \sz \big)
\nonumber\\
\meanemf^{\rm 2s}_y &=&  B_0 \big( (\tilde{\beta}_\perp - \frac{1}{2} \tilde{\mu} ) k_x \cx \sy \sz
     - (\tilde{\delta} + \frac{1}{2} \tilde{\kappa}_\perp) k_y \sx \cy \sz \big)
\nonumber\\
\meanemf^{\rm 2s}_z &=&  - B_0 \big(\tilde{\alpha}_\parallel \sx \sy \sz + \tilde{\kappa}_\parallel k_z \sx \sy \cz \big)
\nonumber
\ENA
and corresponding relations for
$\meanemf_x^{\rm 1c}, \ldots, \meanemf_z^{\rm 2c}$,
whose right-hand sides can be derived from those in \eq{eq057} simply by replacing $\sz$ and $\cz$
by $\cz$ and $-\sz$, respectively.

In view of the assumed axisymmetry of the turbulence,
we consider $\alpha_\perp$, $\alpha_\parallel$, $\ldots$, $\mu$
in what follows as independent of $x$ and $y$ but admit a dependence on $z$.
When multiplying both sides of the equations \eq{eq057} and
of the corresponding ones for
$\meanemf_x^{\rm 1c}, \ldots, \meanemf_z^{\rm 2c}$
with $\sx \sy$, $\sx \cy$ or $\cy \sy$ and averaging over all $x$ and $y$, we obtain a system of equations,
which can be solved for $\tilde{\alpha}_\perp$, $\tilde{\alpha}_\parallel$, $\ldots$, $\tilde{\mu}$.
The result reads
\EQA
\tilde{\alpha}_\perp &=& - \langle b^{ss} (\sz \meanemf^{\rm 1s}_x + \cz \meanemf^{\rm 1c}_x ) \rangle
\nonumber\\
\tilde{\alpha}_\parallel &=& - \langle b^{ss} (\sz \meanemf^{\rm 2s}_z + \cz \meanemf^{\rm 2c}_z ) \rangle
\nonumber\\
\tilde{\gamma} &=& - \langle b^{ss} (\sz \meanemf^{\rm 1s}_y + \cz \meanemf^{\rm 1c}_y ) \rangle
\nonumber\\
\tilde{\beta}_\perp &=& - \half \langle B^{ss} (\cz \meanemf^{\rm 1s}_y - \sz \meanemf^{\rm 1c}_y )
     + B^{sc} (\sz \meanemf^{\rm 2s}_x + \cz \meanemf^{\rm 2c}_x) \rangle
\nonumber\\
&=& - \half \langle B^{ss} (\cz \meanemf^{\rm 1s}_y - \sz \meanemf^{\rm 1c}_y )
     - B^{cs} (\sz \meanemf^{\rm 2s}_y + \cz \meanemf^{\rm 2c}_y) \rangle
\nonumber\\
\tilde{\beta}_\parallel &=& \langle B^{sc} (\sz \meanemf^{\rm 1s}_z + \cz \meanemf^{\rm 1c}_z ) \rangle
\label{eq059}\\
\tilde{\delta} &=& \half \langle B^{ss} (\cz \meanemf^{\rm 1s}_x - \sz \meanemf^{\rm 1c}_x)
     - B^{cs} (\sz \meanemf^{\rm 2s}_x + \cz \meanemf^{\rm 2c}_x) \rangle
\nonumber\\
&=& \half \langle B^{ss} (\cz \meanemf^{\rm 1s}_x - \sz \meanemf^{\rm 1c}_x)
     - B^{sc} (\sz \meanemf^{\rm 2s}_y + \cz \meanemf^{\rm 2c}_y) \rangle
\nonumber\\
\tilde{\kappa}_\perp &=& - \langle B^{ss} (\cz \meanemf^{\rm 1s}_x - \sz \meanemf^{\rm 1c}_x )
     + B^{cs} (\sz \meanemf^{\rm 2s}_x + \cz \meanemf^{\rm 2c}_x ) \rangle
\nonumber\\
&=& - \langle B^{ss} (\cz \meanemf^{\rm 1s}_x - \sz \meanemf^{\rm 1c}_x )
     + B^{sc} (\sz \meanemf^{\rm 2s}_y + \cz \meanemf^{\rm 2c}_y ) \rangle
\nonumber\\
\tilde{\kappa}_\parallel &=& - \langle B^{ss} (\cz \meanemf^{\rm 2s}_z - \sz \meanemf^{\rm 2c}_z) \rangle
\nonumber\\
\tilde{\mu} &=& - \langle B^{ss} (\cz \meanemf^{\rm 1s}_y - \sz \meanemf^{\rm 1c}_y)
     - B^{sc} (\sz \meanemf^{\rm 2s}_x + \cz \meanemf^{\rm 2c}_x) \rangle
\nonumber\\
&=& - \langle B^{ss} (\cz \meanemf^{\rm 1s}_y - \sz \meanemf^{\rm 1c}_y)
     + B^{cs} (\sz \meanemf^{\rm 2s}_y + \cz \meanemf^{\rm 2c}_y) \rangle \, ,
\nonumber
\ENA
where
\EQA
b^{ss} &=& 4 \sx \sy / B_0 \, , \quad B^{ss} = b^{ss} / k_z
\nonumber\\
B^{cs} &=& 4 \cx \sy /k_x B_0 \, , \quad B^{sc} = 4 \sx \cy /k_y B_0.
\label{eq061}
\ENA
The angle brackets indicate averaging over $x$ and $y$.
Although the relations \eq{eq059} and \eq{eq061} contain $k_x$, $k_y$ and $k_z$ as independent variables,
the $\tilde{\alpha}_\perp$, $\tilde{\alpha}_\parallel$, $\ldots$, $\tilde{\mu}$
should vary only via $k_\perp = (k_x^2 + k_y^2)^{1/2}$ with $k_x$ and $k_y$,
and only via $k_\parallel$ with $k_z$.

\subsection{Test-scalar method}

In the passive-scalar case
the coefficients $\gamma^C$, $\beta^C_\perp$, $\beta^C_\parallel$,
and $\delta^C$ are determined by the test-scalar method with
test scalars $\meanC^T$ and the corresponding fluxes $\meanFFFF^T$.
For the latter, we have $\meanFFFF^T=\overline{\uu c^T}$, where $c^T$ obeys
\EQ
{\partial c^T\over\partial t}=-\nab\cdot\left(\meanUU c^T+\uu\meanC^T
+(\uu c^T)'-D\nab c^T\right) \, .
\label{eq071}
\EN
Again $\meanUU$ and $\uu$ are taken from the solutions of \eq{eq041}.

We define two test-scalars $\meanC^{\rm Ts}$ and $\meanC^{\rm Tc}$ by
\EQ
\meanC^{\rm s} = C_0 \sx \sy \sz \, , \quad \meanC^{\rm c} = C_0 \sx \sy \cz \, ,
\label{eq073}
\EN
where $C_0$ is a constant and the abbreviations \eq{eq055} are used.
From \eq{eq029} we then have
\EQA
\meanFFF_x^{\rm s} &=& - C_0 (\tilde{\beta}^C_\perp k_x \cx \sy \sz - \tilde{\delta}^C k_y \sx \cy \sz)
\nonumber\\
\meanFFF_y^{\rm s} &=& - C_0 (\tilde{\beta}^C_\perp k_y \sx \cy \sz + \tilde{\delta}^C k_x \cx \sy \sz)
\label{eq075}\\
\meanFFF_z^{\rm s} &=& - C_0 (\tilde{\gamma}^C_\perp \sx \sy \sz + \tilde{\beta}^C_\parallel k_z \sx \sy \cz)
\nonumber
\ENA
and analogous relations for $\meanFFF_x^{\rm c}, \ldots, \meanFFF_z^{\rm c}$
with $sz$ and $cz$ replaced by $cz$ and $-sz$, respectively.

Analogous to the magnetic case, we assume that $\gamma^C$,
$\beta^C_\perp$, $\beta^C_\parallel$, and $\delta^C$
are independent of $x$ and $y$ but may depend on $z$.
Analogous to \eq{eq059} we find here
\EQA
\tilde{\gamma}^C &=& - \langle c^{ss} (\sz \meanFFF^{\rm s}_z + \cz \meanFFF^{\rm c}_z) \rangle
\nonumber\\
\tilde{\beta}^C_\perp &=& - \langle C^{cs} (\sz \meanFFF^{\rm s}_x + \cz \meanFFF^{\rm c}_x ) \rangle
     = - \langle C^{sc} (\sz \meanFFF^{\rm s}_y + \cz \meanFFF^{\rm c}_y ) \rangle
\nonumber\\
\tilde{\beta}^C_\parallel &=& - \langle C^{ss} (\cz \meanFFF^{\rm s}_z  - \sz \meanFFF^{\rm c}_z ) \rangle
\label{eq077}\\
\tilde{\delta}^C &=& \langle C^{sc} (\sz \meanFFF^{\rm s}_x + \cz \meanFFF^{\rm c}_x) \rangle
     = - \langle C^{cs} (\sz \meanFFF^{\rm s}_y + \cz \meanFFF^{\rm c}_y) \rangle \, ,
\nonumber
\ENA
where $c^{ss}$, $C^{ss}$, $C^{sc}$, and $C^{cs}$ are defined like
$b^{ss}$, $B^{ss}$, $B^{sc}$, and $B^{cs}$,
with $C_0$ at the place of $B_0$.
The angle brackets indicate again averaging over $x$ and $y$.
Note that $\tilde{\gamma}^C$, $\tilde{\beta}^C_\perp$, $\tilde{\beta}^C_\parallel$,
and $\tilde{\delta}^C$ should
depend only via $k_\perp = (k_x^2 + k_y^2)^{1/2}$
on $k_x$ and $k_y$,
and only via $k_\parallel$ on $k_z$.

\subsection{Validation using the Roberts flow}

For a validation of our test-field procedure for the determination
of the coefficients occurring in \eq{eq005} we rely on the Roberts flow.
We define it here by
\EQA
\uu &=& u_0 (- \cos k_0 x \, \sin k_0 y \, , \, \sin k_0 x \, \cos k_0  y \, ,
\nonumber\\
&& \qquad \qquad \qquad \qquad \qquad \qquad
    2 f \cos k_0 x \, \cos k_0 y \,) \, ,
\label{eq081}
\ENA
with some wavenumber $k_0$ and a factor $f$ which characterizes
the ratio of the magnitude of $u_z$ to that of $u_x$ and $u_y$.
We further define mean fields as averages over $x$ and $y$ with an averaging scale
which is much larger than the period length $2 \pi / k_0$ of the flow pattern.
When calculating the mean electromotive force $\meanEMF$ for this flow,
we assume that it is a linear homogeneous function of $\meanBB$
and its first spatial derivatives and adopt the second-order correlation approximation.
Although the Roberts flow is far from being axisymmetric, the result for $\meanEMF$
can be written in the form \eq{eq005}, and we have
\EQA
\alpha_\perp &=& \frac{u_0^2 f}{2 \eta k_0} \, , \quad \alpha_\parallel = \gamma = 0
\nonumber\\
\beta_\perp &=& \frac{u_0^2 (1 + 4 f^2)}{16 \eta k_0^2} \, , \quad
    \beta_\parallel = \frac{u_0^2}{8 \eta k_0^2} \, , \quad \delta = 0
\label{eq083}\\
\kappa_\perp &=& \kappa_\parallel = 0 \, , \quad
\mu = - \frac{u_0^2 (1 - 4 f^2)}{8 \eta k_0^2}
= 2 (\beta_\perp-\beta_\parallel) \, .
\nonumber
\ENA
It agrees with and can be deduced from results reported in \cite{Rae02a,Rae02b}.
As for the passive scalar case, an analogous analytical calculation of the mean scalar flow $\meanFFF$
leads to \eq{eq025} with
\EQ
\gamma^C = 0 \, , \quad \beta^C_\perp = \frac{u_0^2}{8 D k_0^2} \, , \quad
     \beta^C_\parallel = \frac{u_0^2 f^2}{2 D k_0^2} \, , \quad \delta^C = 0 \, .
\label{eq085}
\EN
We may proceed from the local connection of $\meanEMF$ with $\meanBB$ and its derivatives considered in \eq{eq005}
to the non-local ones given by \eq{eq013} or \eq{eq017}.
As a consequence of the deviation of the flow from axisymmetry,
we can then no longer justify that coefficients like $\alpha_\perp (\xxi)$ depend only
via $\xi_x^2 + \xi_y^2$ on $\xi_x$ and $\xi_y$,
and coefficients like $\tilde{\alpha}_\perp (\kk)$ only via $k_\perp$ on $k_x$ and $k_y$.
This applies analogously to the connection of $\meanFFF$ with $\meanC$ and its derivatives
and to coefficients like $\beta_\perp (\xxi)$ and $\tilde{\beta}_\perp (\kk)$.

\begin{figure}[t!]\begin{center}
\includegraphics[width=.9\columnwidth]{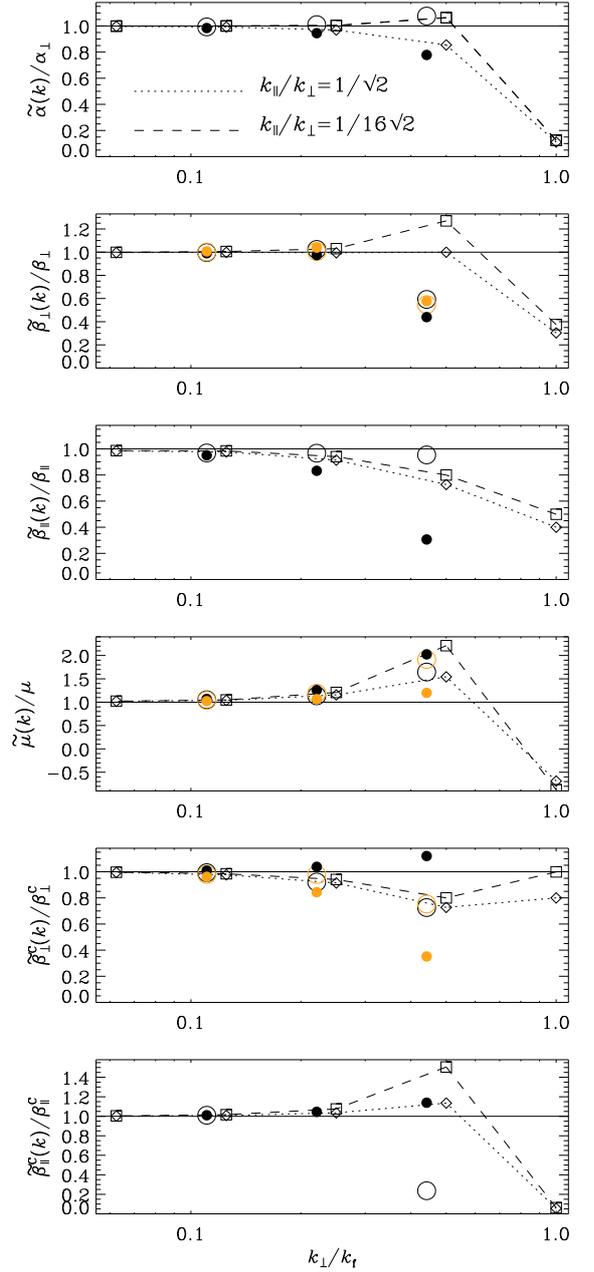}
\end{center}\caption[]{
The coefficients $\tilde{\alpha}_\perp$, $\tilde{\beta}_\perp$, $\tilde{\beta}_\parallel$, and $\tilde{\mu}$,
as well as $\tilde{\beta}_\perp^C$ and $\tilde{\beta}_\parallel^C$ for the Roberts flow,
calculated in the second-order correlation approximation, as functions of $k_\perp/\kf$,
where $\kf=\sqrt{2}k_0$ is the effective wavenumber of the flow.
Results obtained with $k_x = k_y$
and $k_\parallel/k_\perp = 1/\sqrt{2}\approx0.7$ or $k_\parallel/k_\perp = 1/16\sqrt{2}\approx0.004$
are represented by open squares and dotted lines or by open diamonds and dashed lines, respectively.
Results with $k_x/k_y = 0.75$ [$\kk_\perp = (3,4,0) k_1$] or $k_x/k_y = 5$ [$\kk_\perp = (5,1,0) k_1$] and $k_\parallel/k_\perp = 0.2$
are indicated by open or filled circles, respectively.
Orange and black symbols correspond to the first and second expressions for $\tilde{\beta}_\perp$
and $\tilde{\mu}$ in \eq{eq059}
or for $\tilde{\beta}_\parallel^C$ in \eq{eq077}.
}\label{proberts}\end{figure}

A test-field calculation of the coefficients $\tilde{\alpha}_\perp$, $\tilde{\alpha}_\parallel$, $\ldots$, $\tilde{\mu}$,
as well as $\tilde{\gamma}^C$, $\ldots$, $\tilde{\delta}^C$,
has been carried out under the conditions of the second-order correlation approximation
with $\uu$ given by \eq{eq081} and $f = 1 / \sqrt{2}$.
\Fig{proberts} shows the results obtained for $\tilde{\alpha}_\perp$, $\tilde{\beta}_\perp$,
$\tilde{\beta}_\parallel$ and $\tilde{\mu}$,
as well as $\tilde{\beta}_\perp^C$ and $\tilde{\beta}_\parallel^C$,
as functions of $k_\perp/\kf$, with $\kf=\sqrt{2}k_0$, for two fixed ratios $k_\parallel / k_\perp$.
In the limit $k_\perp/\kf\ll1$ these coefficients take just the values of $\alpha_\perp$, $\beta_\perp$,
$\beta_\parallel$, $\mu$, $\beta^C_\perp$ and $\beta^C_\parallel$ given in \eq{eq083} and \eq{eq085}.
For larger values of $k_\perp/\kf$, as to be expected, the $\tilde{\alpha}_\perp$, $\tilde{\beta}_\perp$,
$\tilde{\beta}_\parallel$, $\tilde{\mu}$, $\tilde{\beta}_\perp^C$ and $\tilde{\beta}_\parallel^C$
depend also on the ratio of $k_x$ and $k_y$.

\subsection{Dimensionless parameters and related issues}

Within the framework of this paper, the coefficients $\alpha_\perp$,
$\alpha_\parallel$, $\ldots$, $\mu$ as well as
$\tilde{\alpha}_\perp$, $\tilde{\alpha}_\parallel$, $\ldots$, $\tilde{\mu}$,
and likewise $\gamma^C$, $\beta^C_\perp$, $\ldots$, $\delta^C$
and $\tilde{\gamma}^C$, $\tilde{\beta}^C_\perp$, $\ldots$, $\tilde{\delta}^C$,
have to be considered as functions of several dimensionless parameters.
In the magnetic case these are the magnetic Reynolds number $\Rm=\urms/\eta\kf$
and the magnetic Prandtl number $\Pm=\nu/\eta$,
in the passive scalar case the P\'eclet number $\Pe=\urms/D\kf$
and the Schmidt number $\Sc=\nu/D$,
further
the Mach number $\Ma=\urms/\cs$, the gravity parameter $\Gr=g/\cs^2\kf$,
the Coriolis number $\Co = 2\Omega/\urms\kf$,
as well as the scale separation ratio $\kf/k_1$.

Throughout the rest of the paper we give the coefficients $\alpha_\perp$, $\alpha_\parallel$, $\gamma$,
and $\gamma^C$ as well as $\tilde{\alpha}_\perp$, $\tilde{\alpha}_\parallel$, $\tilde{\gamma}$,
and $\tilde{\gamma}^C$ in units of $\urms/3$,
the remaining coefficients $\beta_\perp$, $\ldots$,
$\delta^C$ and $\tilde{\beta}_\perp$, $\ldots$, $\tilde{\delta}^C$
in units of $\urms/3\kf$.
The numerical calculations deliver these coefficients as functions of
$z$ and $t$.
To avoid boundary effects, we average these results over
$-2\le k_1z\leq1$ (see \Fig{ppanisoz} below).
The resulting time series are averaged over a range where the results
are statistically stationary, i.e., there is no trend in the time series.
Error bars are defined by comparing the maximum departure of an average
over any one third of the time series with the full time average.

In the case of isotropic turbulence it has been observed that
many of the transport coefficients enter an asymptotic regime
as soon as $\Rm$ exceeds unity \citep{SBS08}.
While this should be checked in every new case again (see below),
it is important to realize that, according to several earlier results
\citep[see also][]{BSV09}, only values of $\Rm$ below unity
are characteristic of the diffusively dominated regime, while for
$\Rm$ exceeding unity
the transport coefficients turn out to be nearly independent of the value of $\Rm$.

We are often interested in the limit $k_\perp, k_\parallel \to 0$,
in which $\tilde{\alpha}_\perp$, $\tilde{\alpha}_\parallel$, $\ldots$ $\tilde{\delta}^C$
turn into $\alpha_\perp$, $\alpha_\parallel$ $\ldots$ $\delta^C$.
In this limit, however, the test fields and test scalars defined by \eq{eq053} and \eq{eq073} vanish.
Unless specified otherwise, we approach this limit by choosing the smallest possible non-zero $|k_x|$, $|k_y|$ and $|k_z|$,
that is, by putting $k_x=k_y=k_z=k_1$.

In the figures of the next section results for
$\tilde{\alpha}_\perp$, $\tilde{\alpha}_\parallel$, $\ldots$ $\tilde{\delta}^C$ are represented.
In all cases in which they are considered as results for the limit $k_\perp, k_\parallel \to 0$
they are simply denoted as $\alpha_\perp$, $\alpha_\parallel$ $\ldots$ $\delta^C$ in the text.

\section{Results}

\subsection{Homogeneous rotating turbulence}

Let us first consider homogeneous turbulence in a rotating system,
that is, under the influence of the Coriolis force.
The angular velocity $\OO$ responsible for this force defines the preferred direction
of the turbulence, $\eee = \OO / |\OO|$.
In this case we expect only contributions to the mean electromotive force $\meanEMF$
from a spatially varying mean magnetic field $\meanBB$,
and contributions to the passive scalar flux $\meanFFFF$ from
a spatially varying mean passive scalar $\meanC$.
That is, in \eq{eq005} we have only the
terms with $\beta_\perp$, $\beta_\parallel$, $\delta$,
$\kappa_\perp$, $\kappa_\parallel$, and $\mu$,
and in \eq{eq025} only those with $\beta_\perp^C$, $\beta_\parallel^C$, and $\delta^C$.
The terms with $\beta_\perp$ and $\beta_\parallel$,
as well as those with $\beta_\perp^C$ and $\beta_\parallel^C$,
characterize anisotropic mean-field diffusivities, and
that with $\delta$ corresponds to the ``$\OO \times \meanJJ$ effect''
\citep{Rae69a,Rae69b,Rae76,KR71,KR80,RKR03}, while the $\delta^C$ term
vanishes underneath the divergence and is therefore without interest.

\begin{figure}[t!]\begin{center}
\includegraphics[width=\columnwidth]{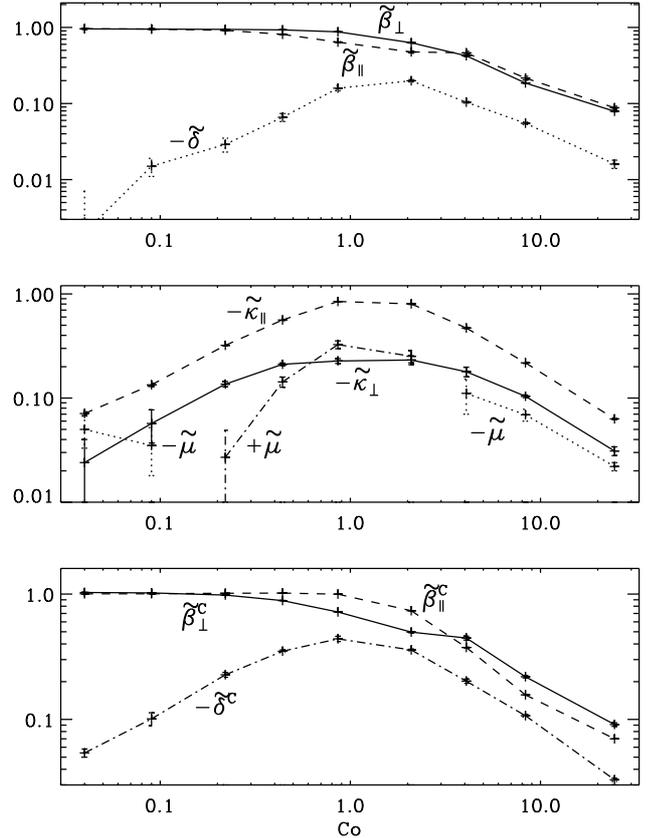}
\end{center}\caption[]{
$\Co$ dependence of transport coefficients in a model with rotation
but zero
density stratification, $\Rm\approx9$, $\Pm=\Sc=1$, $\Gr=0$, $\kf/k_1=5$.
}\label{psummary_Odep_g0}\end{figure}

\FFig{psummary_Odep_g0} shows the dependence of the aforementioned
coefficients on $\Co$ for $\Rm\approx\Pe\approx9$ and $\kf/k_1=5$.
The values of $\beta_\perp$, $\beta_\parallel$, $\beta_\perp^C$ and $\beta_\parallel^C$,
which remain finite for $\Co \to 0$, are always close together.
The other four coefficients vary linearly with $\Co$ as long as $\Co$ is small.
Specifically, we find $\tilde\delta\approx-0.1\,\Co$,
$\tilde\delta^C\approx-\Co$, as well as
$\tilde\kappa_\perp\approx-0.3\,\Co$ and $\tilde\kappa_\parallel\approx-\Co$.
These coefficients reach maxima at $\Co\approx1$.
For rapid rotation, $|\Co|\gg1$, all coefficients approach zero like $1/\Co$.
In particular, we have
$\beta_\perp\approx1.2/\Co$ and the same for $\beta_\parallel$,
$\beta_\perp^C$, and $\beta_\parallel^C$, further
$\tilde\kappa_\perp\approx-0.5/\Co$, $\tilde\kappa_\parallel\approx-1.2/\Co$,
$\tilde\delta\approx-0.3/\Co$, and $\tilde\delta^C\approx-0.6/\Co$.
Furthermore, we find that, within error bars, $\alpha_\perp$,
$\alpha_\parallel$, $\gamma$, and $\gamma^C$ are indeed zero.

\subsection{Stratified turbulence}

\begin{figure}[t!]\begin{center}
\includegraphics[width=\columnwidth]{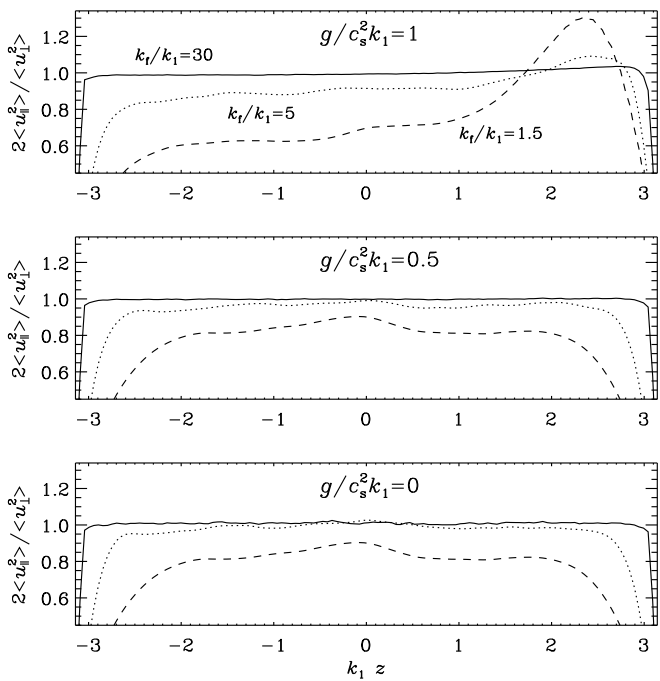}
\end{center}\caption[]{
Anisotropy $2 \overline{u_\parallel^2}/\overline{u_\perp^2}$ of
nonrotating turbulence
for different stratifications, $g/\cs^2 k_1$,
and different degrees of scale separation, $\kf/k_1$.
}\label{ppanisoz}\end{figure}

Owing to the presence of boundary conditions at the top and bottom of our domain
and the lack of scale separation for our default choice of $\kf/k_1=5$,
the turbulence is in all cases anisotropic, even if gravity is negligible.
The ratio of the vertical and horizontal velocity components, $2 \overline{u_\parallel^2}/\overline{u_\perp^2}$,
is no longer, as in the isotropic case, equal to unity.
For moderate stratification ($g/\cs^2 k_1\approx1$),
not too large $|z|$, and $\kf/k_1=5$, it takes a value of about $0.9$.
It decreases when the ratio $\kf/k_1$ is decreased; see \Tab{paniso}.
\Fig{ppanisoz} shows the $z$ dependence of $2 \overline{u_\parallel^2}/\overline{u_\perp^2}$.
For strong stratification and a high degree of scale separation, e.g.\ $\kf/k_1=30$,
the mentioned ratio comes close to unity.
Note, however, that smaller values of
$2\overline{u_\parallel^2}/\overline{u_\perp^2}$
can be can be achieved in the non-isothermal case
when the effects of buoyancy become important.

\begin{table}[b!]\caption{
Dependence of the density contrast $\rho_{\rm bot}/\rho_{\rm top}$
and the degree of anisotropy $2 \overline{u_\parallel^2}/\overline{u_\perp^2}$,
for three different values of $\kf/k_1$,
on the density stratification $g/\cs^2 k_1$ for nonrotating turbulence.
The values of $2 \overline{u_\parallel^2}/\overline{u_\perp^2}$
have been obtained as averages over the range $-2\leq k_1 z\leq1$.
}\vspace{12pt}\centerline{\begin{tabular}{c|c|ccc}
\hline
\vspace{-3mm}
\\
$\!g/\cs^2 k_1$ & $\rho_{\rm bot}/\rho_{\rm top}$
 & \multicolumn{3}{|c}{$2 \overline{u_\parallel^2}/\overline{u_\perp^2}$} \\
 & &
$\kf=1.5k_1\!$ & $\!\kf=5k_1\!$ & $\!\kf=30k_1\!$ \\
\hline
 0  &   0 & 0.84 & 0.99 & 1.00  \\
0.5 &  23 & 0.84 & 0.97 & 1.00 \\
 1  & 540 & 0.66 & 0.90 & $\;$0.99
\label{paniso}\end{tabular}}\end{table}

\subsubsection{Stratified nonrotating turbulence}

\begin{figure}[t!]\begin{center}
\includegraphics[width=\columnwidth]{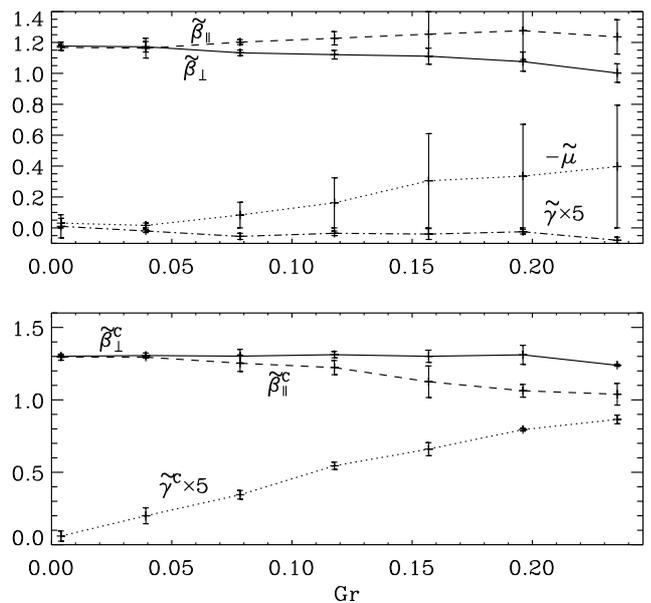}
\end{center}\caption[]{
$\Gr$ dependence of the transport coefficients in a model with density stratification
but zero rotation,
$\Pm=\Sc=1$, $\Rm\approx22$, $\Co=0$, $\kf/k_1=5$.
}\label{psummary_gdep}\end{figure}

For axisymmetric turbulence in a nonrotating system showing any kind of stratification
in the representation \eq{eq005} of $\meanEMF$ only the four coefficients
$\gamma$, $\beta_\perp$, $\beta_\parallel$, and $\mu$ can be non-zero.
Likewise, in the representation \eq{eq025} of $\meanFFFF$ only the three coefficients
$\gamma^C$, $\beta_\perp^C$, and $\beta_\parallel^C$ can be non-zero.
\Fig{psummary_gdep} shows their dependence on $\Gr$.
It appears that $\gamma$ is always close to zero,
while $\gamma^C$ shows a linear increase for not too strong gravity.
At the same time, $\beta_\perp$, $\beta_\parallel$,
$\beta_\perp^C$, and $\beta_\parallel^C$ remain approximately constant.
We find that $\mu$ is negative and its modulus is
mildly increasing with increasing
stratification, but the error bars are large.

\subsubsection{Stratified rotating turbulence}

For turbulence under the influence of gravity and rotation, all nine
coefficients $\alpha_\perp$, $\ldots$, $\mu$ are in general non-zero,
as well as all four coefficients $\gamma^C$, $\ldots$, $\delta^C$.
If both gravity and rotation are so small that $\meanEMF$ is linear in $g$ and $\Omega$,
more precisely $\meanEMF$ contains $g^m \Omega^n$, where $n$ and $m$ mean integers, only with $n+m \leq 1$,
$\alpha_\perp$ and $\alpha_\parallel$ vanish but $\gamma$, $\beta_\perp$, $\delta$ and $\kappa_\perp$
may well be unequal to zero.
If $n+m \leq 2$, all nine coefficients may indeed be non-zero.

\begin{figure}[t!]\begin{center}
\includegraphics[width=\columnwidth]{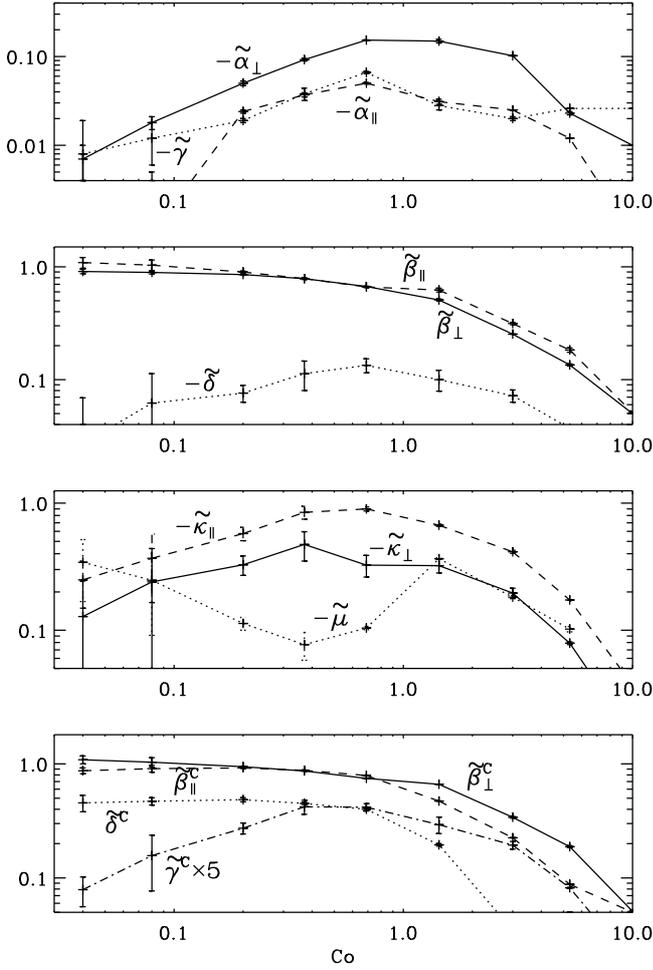}
\end{center}\caption[]{
$\Co$ dependence of transport coefficients in a model with rotation and
density stratification, $\Pm=\Sc=1$, $\Rm\approx10$, $\Gr\approx0.16$, $\kf/k_1=5$.
}\label{psummary_Odep_g08}\end{figure}

Results for stratified rotating turbulence are
shown in \Fig{psummary_Odep_g08}.
The error bars are now bigger than either with just rotation or
just stratification.
For $\Co \to 0$,
the coefficients $\beta_\perp$, $\beta_\parallel$, $\mu$, $\beta^C_\perp$, $\beta^C_\parallel$ and $\delta_C$
remain finite.
As $\Co$ is increased,
their moduli show some decline.
On the other hand
the moduli of $\alpha_\perp$, $\alpha_\parallel$, $\gamma$, $\delta$, $\kappa_\perp$, $\kappa_\parallel$
and $\gamma^C$ increase with $\Co$
as long as it is smaller than some value below unity but decrease again for larger $\Co$.
Both $\alpha_\perp$ and $\alpha_\parallel$ are negative,
which is expected for $\grav$ and $\OO$ being antiparallel to each other.
Interestingly, $\mu$ is finite for small values of $\Co$, in agreement
with the result when there is only stratification (\Fig{psummary_gdep}),
but with a modest amount of rotation, $\mu$ is suppressed and grows
only when $\Co$ has reached values around unity.

\subsection{Wavenumber dependence}

\begin{figure}[t!]\begin{center}
\includegraphics[width=\columnwidth]{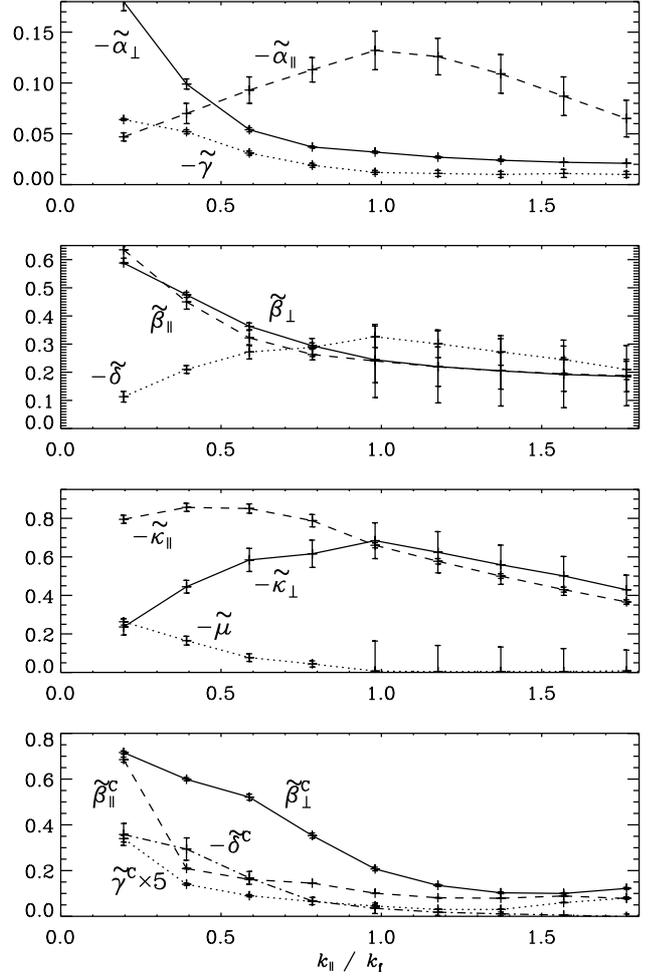}
\end{center}\caption[]{
$k_\parallel$ dependence of transport coefficients in a model with rotation
and density stratification,
$k_\perp = \sqrt{2}k_1$,
$\Pm = \Sc = 1$,
$\Rm=12$, $\Co=1.0$,
$\Gr = 0.16$,
$\kf/k_1=5$.
}\label{psummary_kdep}\end{figure}

So far we have considered the coefficients $\tilde{\alpha}_\perp$, $\tilde{\alpha}_\perp$, $\ldots$,
$\tilde{\delta}^C$ in the limit $k = |\kk| \to 0$, that is, $k_\perp, k_\parallel \to 0$.
However, their behavior for larger $k$, in particular for $k$ up to several $\kf$, is of interest, too.
Most of them decrease like $k^{-2}$ as $k$ grows and can
be fitted to a Lorentzian profile, as has been found in earlier
calculation using the test-field method; see \cite{BRS08},
where in fact the dependence on $k_\parallel$ was considered.
Even earlier work that was not based on the test-field method
showed a declining trend \citep{MBZ00,BS02}.
Nevertheless, as is shown in \Fig{psummary_kdep},
there are also some coefficients
that first increase with $k_\parallel$,
have a maximum near
$k_\parallel = \kf$
and only then decrease with growing $k_\parallel$.
Examples for such a behavior are
$\tilde{\alpha}_\parallel$, $\tilde{\delta}$, and $\tilde{\kappa}_\perp$,
while $\tilde{\kappa}_\parallel$ peaks slightly below
$k_\parallel = 0.5 \kf$.

\begin{figure}[t!]\begin{center}
\includegraphics[width=\columnwidth]{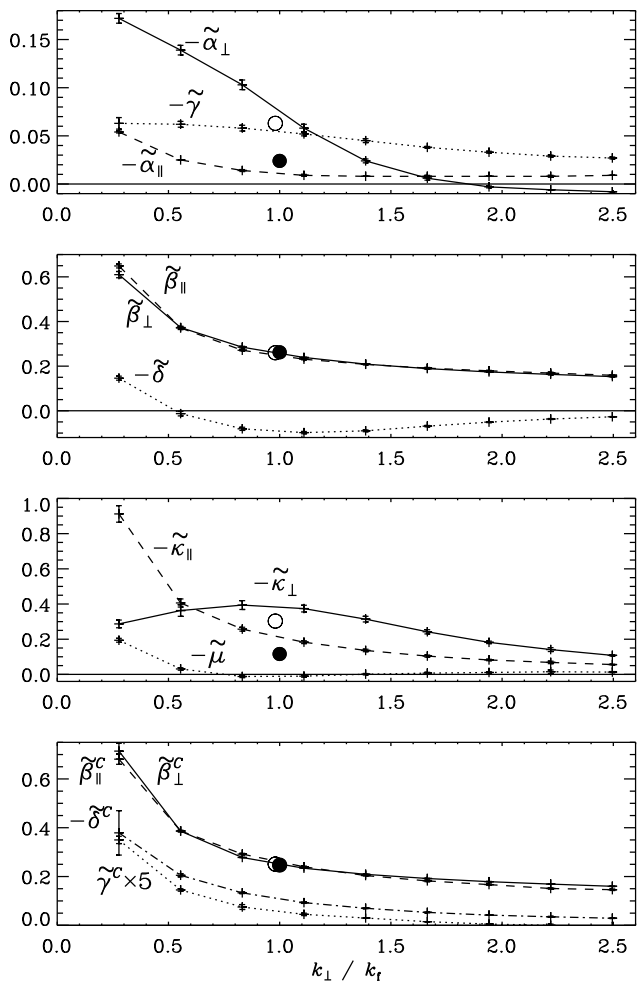}
\end{center}\caption[]{
Same as \Fig{psummary_kdep}, but $k_\perp$ dependence,
$k_\parallel = k_1$.
The filled and open circles denote results for $\alpha_\perp$,
$\beta_\perp$, $\kappa_\perp$, and $\beta_\perp^C$ obtained with $k_x/k_y = 0.75$ [$\kk_\perp=(3,4,0)k_1$]
and $k_x/k_y = 0.2$ [$\kk_\perp=(1,5,0)k_1$], respectively.
}\label{psummary_kperpdep}\end{figure}

The dependence of the coefficients under discussion
on $k_\perp$ is shown in \Fig{psummary_kperpdep}.
Note that our test fields vanish for $k_\perp=0$, so no values
are shown for this case.
Note also that $-\tilde{\alpha}_\parallel$, $-\tilde{\delta}$,
and $-\tilde{\kappa}_\parallel$, which have maxima for
$k_\parallel / k_{\rm f} \approx 1$ or $k_\parallel / k_{\rm f} \approx 0.5$,
show a clear monotonic decline with $k_\perp$.
Only $-\tilde{\kappa}_\perp$
has maxima with respect to $k_\parallel/\kf$ and $k_\perp/\kf$.

Most of the results presented in \Fig{psummary_kperpdep} have been calculated with $k_x = k_y$,
a few single ones for $\tilde{\alpha}_\perp$, $\tilde{\beta}_\perp$, $\tilde{\kappa}_\perp$ and $\tilde{\beta}^C_\perp$
also with $k_x/k_y = 0.75$ and $k_x/k_y = 0.2$.
While the results for  $\tilde{\beta}_\perp$ and $\tilde{\beta}^C_\perp$ agree well for all these values of $k_x/k_y$,
there are significant
discrepancies with $\tilde{\alpha}_\perp$ and $\tilde{\kappa}_\perp$.

\subsection{Dependencies on $\Rm$ and $\Pe$}

Let us finally consider the dependence of all transport coefficients
on $\Rm$ or $\Pe$ for a case where they are all expected to be finite.
Therefore we choose again the case
with $\Co=1$ and $\Gr=0.16$,
which was also
considered in \Figss{psummary_Odep_g08}{psummary_kperpdep},
and keep
$\Pm=\Sc=1$.

\begin{figure}[t!]\begin{center}
\includegraphics[width=\columnwidth]{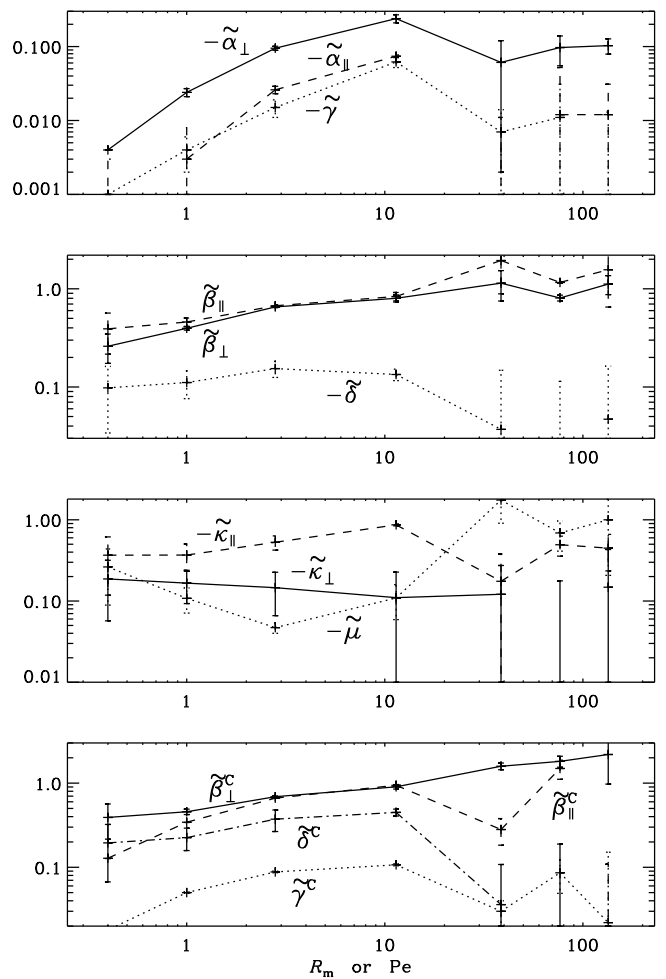}
\end{center}\caption[]{
Dependencies of the transport coefficients on $\Rm$ or $\Pe$
in a model with rotation and density stratification,
$\Pm = \Sc = 1$,
$\Co=1.0$, $\Gr=0.16$,
$\kf/k_1=5$.
}\label{psummary_Rmdep}\end{figure}

The results are shown in \Fig{psummary_Rmdep}.
As expected, some of the quantities increase approximately linearly
with $\Rm$ if $\Rm < 1$, or with $\Pe$ if $\Pe < 1$,
and seem to level off to constant values for larger values of $\Rm$,
or $\Pm$, although the uncertainty tends to increase significantly.

\section{Conclusions}

In this paper we have dealt with the mean electromotive force and the mean passive scalar flux
in axisymmetric turbulence and have calculated
the transport coefficients that define these quantities.
Unlike most of the earlier work, we have no longer assumed
that mean fields are defined as planar averages but admit a dependence
on all three space coordinates.
The number of test fields and test scalars is the same
(4 and 2, respectively) as in earlier work using planar averages,
so the computational cost is unchanged.

We may conclude from general symmetry considerations that the mean electromotive force $\meanEMF$
has altogether nine contributions:
three defined by the mean magnetic field $\meanBB$,
three by the mean current density $\meanJJ$,
and three by the vector
$\meanKK$, which is the projection of the symmetric part
of the gradient tensor $\nab \meanBB$ of the magnetic field
on the preferred direction.
In many representations of $\meanEMF$ the last three contributions have been ignored.
Our results underline that this simplification is in general not justified.
The corresponding coefficients $\kappa_\perp$, $\kappa_\parallel$ and $\mu$ are in general not small
compared to $\beta_\perp$, $\beta_\parallel$ and $\delta$.

It has been known since long that a stratification of the turbulence intensity,
that is, a gradient of $\overline{\uu^2}$, causes a pumping of magnetic
flux \citep{Rae66,Rae68,Rae69b}.
It remained however uncertain whether the same effect occurs if a preferred direction is given
by a gradient of the mean mass density $\overline{\rho}$ while the turbulence intensity is spatially constant.
In our calculations, which correspond to this assumption, the value of $\gamma$ is not clearly different from zero.
This suggests that a gradient of the mass density alone is not sufficient for pumping,
what is also in agreement with results of \cite{BKKR11}.
This is even more remarkable as the corresponding coefficient $\gamma^C$  which describes the transport
of a mean passive scalar is noticeably different from zero.
Pumping down the density gradient is indeed expected \citep{EKR95}.
An explanation of these results would be very desirable.

In homogeneous rotating turbulence, apart from an anisotropy of the mean-field conductivity,
the $\OO\times\meanJJ$ effect occurs \citep{Rae69a,Rae69b}.
In the passive scalar case again an anisotropy of the mean diffusivity is possible.
Even if the flux proportional to $\OO \times \nab \meanC$ is non-zero, it cannot influence $\meanC$.

Let us turn to the induction effects described by $\meanKK$.
If the preferred direction is given by a polar vector, the corresponding contribution
to the mean electromotive force can only be proportional to $\eee \times \meanKK$.
We found such a contribution in the case of the Roberts flow
and also, for turbulence subject the Coriolis force, in the results presented in
\Fig{psummary_Odep_g0} and \Figss{psummary_gdep}{psummary_kperpdep}.

Contributions to the mean electromotive force as
described here by $\meanKK$ occur also in earlier calculations,
e.g.\ \cite{KPR94} or \cite{RB95}.
As a consequence of other notations, however, this is not always obvious.
For example, \cite{RB95} consider a mean electromotive force of the form
\EQ
\meanEMF = - \eta_\parallel \meanJJ + (\eta_\parallel - \eta_{\rm T}) (\zzz \meanJ_z - \zzz \times \nab \meanB_z)
\label{eq101}
\EN
with two coefficients $\eta_\parallel$ and $\eta_{\rm T}$
(equation (18) of their paper with $\mu_0 \meanJJ$, in the sense of the definition introduced here,
replaced by $\meanJJ$; $\zzz$ is our $\eee$).
It is equivalent to our representations \eq{eq005} or \eq{eq007} of $\meanEMF$
if we put there $\beta_\perp = \half (\eta_\parallel + \eta_{\rm T})$, $\beta_\parallel = \eta_{\rm T}$,
$\mu = \eta_\parallel - \eta_{\rm T}$
and all other coefficients equal to zero.
This implies $\beta_\perp - \beta_\parallel = \mu/2$,
which is in agreement with the relation for $\mu$ in equation \eq{eq083} for the Roberts flow.
The latter equality is also approximately obeyed for turbulence in the
presence of rotation, stratification, and both; see
Figs.~\ref{psummary_Odep_g0}, \ref{psummary_gdep}, and \ref{psummary_Odep_g08},
respectively.

If there is moderate rotation ($\Co\approx1$), but no stratification,
we have $\beta_\perp > \beta_\parallel$; see \Fig{psummary_Odep_g0}.
This means, e.g., that for a magnetic field without a component in the direction
of the rotation axis the diffusion along this axis is enhanced compared
with that in the perpendicular direction.
In the passive scalar case we have $\beta^C_\parallel>\beta^C_\perp$,
which implies that the diffusion along the rotation axis is enhanced, too.
However, stratification enlarges $\beta_\parallel - \beta_\perp$
and diminishes $\beta^C_\parallel - \beta^C_\perp$ so that the diffusion
along the rotational axis is decreased in both cases considered.
In the presence of rotation and density stratification
all three contributions to the mean electromotive force described by $\meanKK$
are in general non-zero.
Here, $|\kappa_\perp|$ is smaller than $|\kappa_\parallel|$.
There is now also an $\alpha$ effect, which is necessarily anisotropic,
and $|\alpha_\parallel|$ is typically only half as big as $|\alpha_\perp|$;
see \Fig{psummary_Odep_g08}.

The present work is applicable to investigations of stellar convection
either with or without rotation, and it would provide a more comprehensive
description of turbulent transport properties than what has been
available so far \citep{KKB09}.
The methods
utilized in this paper can be extended to a large class of phenomena
in which turbulence with just one preferred direction plays an important role.
Examples for that include turbulence under the influence of a strong magnetic field
and/or an externally applied electric field leading to a current permeating the system.
Turbulence generated by the \cite{Bell04} instability is an example.
In addition to density stratification, there can be a systematic
variation of the turbulence intensity in one direction.
A further example is entropy inhomogeneity combined with gravity giving
rise to Brunt-V\"ais\"al\"a oscillations.
Pumping effects also exist in homogeneous flows if the turbulence
is helical \citep{Mitra09,RKKB11}.
By contrast, shear problems or other types of problems with two or more preferred
directions that are inclined to each other
(e.g., turbulence in a local domain of a rotating stratified shell at latitudes different
from the two poles) are not amenable to such a study.
Of course, although we refer here to axisymmetric turbulence, problems in
axisymmetric cylindrical geometry are also not amenable to this method,
because the turbulence must be homogeneous in one plane.

\begin{acknowledgements}
A.B.\ and K.-H.R.\ are grateful for the opportunity to work on this paper
while participating in the program ``The Nature of Turbulence"
at the Kavli Institute for Theoretical Physics in Santa Barbara, CA.
This work was supported in part by
the European Research Council under the AstroDyn Research Project No.\ 227952
and by the National Science Foundation under Grant No.\ NSF PHY05-51164.
We acknowledge the allocation of computing resources provided by the
Swedish National Allocations Committee at the Center for
Parallel Computers at the Royal Institute of Technology in
Stockholm and the National Supercomputer Centers in Link\"oping.
\end{acknowledgements}

\appendix

\section{Derivation of relation \eq{eq005}}
\label{app1}

We start from the aforementioned assumption according to which $\meanEMF$
is linear and homogeneous in $\meanBB$ and its first spatial derivatives,
\EQ
\meanemf_i  = a_{ij} \meanB_j + b_{ijk} (\nab \meanBB)_{jk} \, .
\label{eqA01}
\EN
Here $a_{ij}$ and $b_{ijk}$ are tensors determined by the fluid flow.
The gradient tensor $(\nab \meanBB)_{jk}$ can be split into an
antisymmetric part, which can be expressed by $\meanJJ$,
and a symmetric part $(\nab \meanBB)^{\rm{S}}_{jk}$.
Therefore we may also write
\EQ
\meanemf_i  = a_{ij} \meanB_j - b_{ij} \meanJ_j - c_{ijk} (\nab \meanBB)^{\rm{S}}_{jk}
\label{eqA03}
\EN
with new tensors $b_{ij}$ and $c_{ijk}$,
the latter being symmetric in $j$ and $k$.
From the further assumption that the flow constitutes an axisymmetric turbulence
we may conclude that $a_{ij}$, $b_{ij}$ and $c_{ijjk}$ are axisymmetric tensors.
Defining the preferred direction by the unit vector $\eee$ we then have
\EQA
a_{ij} &=& a_1 \delta_{ij} + a_2 \epsilon_{ijl} \hat{e}_l + a_3 \hat{e}_i \hat{e}_j \, ,
\nonumber\\
b_{ij} &=& b_1 \delta_{ij} + b_2 \epsilon_{ijl} \hat{e}_l + b_3 \hat{e}_i \hat{e}_j \, ,
\nonumber\\
c_{ijk} &=& c_1 \delta_{jk} \hat{e}_i + c_2 (\delta_{ij} \hat{e}_k + \delta_{ik} \hat{e}_j)
\label{eqA05}\\
&& + c_3 (\epsilon_{ijl} \hat{e}_l \hat{e}_k + \epsilon_{ikl} \hat{e}_l \hat{e}_j)
    + c_4 \hat{e}_i \hat{e}_j \hat{e}_k \, ,
\nonumber
\ENA
with coefficients $a_1$, $a_2$, $\ldots$, $c_4$ determined by the fluid flow.
Taking \eq{eqA03} and \eq{eqA05} together and considering that
\EQA
(\delta_{ij} \hat{e}_k + \delta_{ik} \hat{e}_j) (\nab \meanBB)^{\rm{S}}_{jk} &=& 2 \overline{K}_i \, ,
\nonumber\\
(\epsilon_{ijl} \hat{e}_l \hat{e}_k + \epsilon_{ikl} \hat{e}_l \hat{e}_j) (\nab \meanBB)^{\rm{S}}_{jk}
    &=& - 2 (\eee \times \meanKK)_i \, ,
\label{eqA07}\\
\hat{e}_i \hat{e}_j \hat{e}_k (\nab \meanBB)^{\rm{S}}_{jk} &=&  (\eee \cdot \meanKK) \hat{e}_i \, ,
\nonumber
\ENA
we find
\EQA
\meanEMF &=& a_1 \meanBB - a_2 \eee \times \meanBB - a_3 (\eee \cdot \meanBB) \eee
\nonumber\\
&& + b_1 \meanJJ - b_2 \eee \times \meanJJ - b_3 (\eee \cdot \meanJJ) \eee
\label{eqA09}\\
&& + 2 c_2 \meanKK - 2 c_3 \eee \times \meanKK  + c_4 (\eee \cdot \meanKK) \eee \, .
\nonumber
\ENA
Since $(\nab \meanBB)_{ii} = 0$ there is no contribution with $c_1$.
With a proper renaming of the coefficients \eq{eqA09} turns into \eq{eq005}.

\newcommand{\ymonber}[3]{ #1, {Monats.\ Dt.\ Akad.\ Wiss.,} {#2}, #3}

\vfill\bigskip\noindent\tiny\begin{verbatim}
$Header: /var/cvs/brandenb/tex/koen/rotstrat/paper.tex,v 1.168 2012-02-05 12:21:04 brandenb Exp $
\end{verbatim}


\begin{thebibliography}{}

\bibitem[Bell(2004)]{Bell04}
Bell, A. R.\ymn{2004}{353}{550}

\bibitem[Brandenburg \& Sokoloff(2002)]{BS02}
Brandenburg, A., \& Sokoloff, D.\ygafd{2002}{96}{319}

\bibitem[Brandenburg \& Subramanian(2005)]{BS05}
Brandenburg, A., \& Subramanian, K.\yjour{2005}{Phys.\ Rep.}{417}{1}

\bibitem[Brandenburg et al.(2008a)]{BRS08}
Brandenburg, A., R\"adler, K.-H., \& Schrinner, M.\yana{2008a}{482}{739}

\bibitem[Brandenburg et al.(2008b)]{BRRK08}
Brandenburg, A., R\"adler, K.-H., Rheinhardt, M., \& K\"apyl\"a, P. J.\yapj{2008b}{676}{740}

\bibitem[Brandenburg et al.(2009)]{BSV09}
Brandenburg, A., Svedin, A., \& Vasil, G. M.\ymn{2009}{395}{1599}

\bibitem[Brandenburg et al.(2011)]{BKKR11}
Brandenburg, A., Kemel, K., Kleeorin, N., \& Rogachevskii, I.\sapj{2011}

\bibitem[Chatterjee et al.(2011)]{CMRB11}
Chatterjee, P., Mitra, D., Rheinhardt, \& M. Brandenburg, A.\yana{2011}{534}{A46}

\bibitem[Elperin et al.(1995)]{EKR95}
Elperin, T., Kleeorin, N., Rogachevskii, I.\yprl{1995}{52}{2617}

\bibitem[Elperin et al.(1996)]{EKR96}
Elperin, T., Kleeorin, N., Rogachevskii, I.\yprl{1996}{76}{224}

\bibitem[Haugen et al.(2004)]{HBD04}
Haugen, N.E.L., Brandenburg, A., Dobler, W.\ypre{2004}{70}{016308}

\bibitem[Hubbard \& Brandenburg(2009)]{HB09}
Hubbard, A., \& Brandenburg, A.\yapj{2009}{706}{712}

\bibitem[K\"apyl\"a et al.(2009)]{KKB09}
K\"apyl\"a, P. J., Korpi, M. J., \& Brandenburg, A.\yana{2009}{500}{633}

\bibitem[Kitchatinov et al.(1994)]{KPR94}
Kitchatinov, L. L., R\"udiger, G., Pipin, V. V. \& R\"udiger, G.\yan{1994}{315}{157}

\bibitem[Krause \& R\"adler(1971)]{KR71}
Krause, F., \& R\"adler, K.-H., 1971, In R. Rompe and M. Steenbeck,
{\it Ergebnisse der Plasmaphysik und der Gaselektronik} Band 2,
Akademie-Verlag Berlin, pp.\ 6--154

\bibitem[Krause \& R\"adler(1980)]{KR80}
Krause, F., \& R\"adler, K.-H., 1980,
{\it Mean-Field Magneto\-hydro\-dynamics and Dynamo Theory},
Akademie-Verlag Berlin and Pergamon Press Cambridge

\bibitem[Madarassy \& Brandenburg(2010)]{MB10}
Madarassy, E. J. M., \& Brandenburg, A.\ypre{2010}{82}{016304}

\bibitem[Miesch et al.(2000)]{MBZ00}
Miesch, M. S., Brandenburg, A., \& Zweibel, E. G.\ypr{2000}{E61}{457}

\bibitem[Mitra et al.(2009)]{Mitra09}
Mitra, D., K\"apyl\"a, P. J., Tavakol, R., \& Brandenburg, A.\yana{2009}{495}{1}

\bibitem[Prandtl(1925)]{Pra25}
Prandtl, L.\yjour{1925}{Zeitschr.\ Angewandt.\ Math.\ Mech.}{5}{136}

\bibitem[R\"adler(1966)]{Rae66}
R\"adler, K.-H. 1966, Thesis Univ. Jena

\bibitem[R\"adler(1968)]{Rae68}
R\"adler, K.-H. 1968, Z.Naturforschg. 23a, 1851

\bibitem[R\"adler(1969a)]{Rae69a}
R\"adler, K.-H. 1969a, Mber. Dt. Akad. Wiss 11, 194

\bibitem[R\"adler(1969b)]{Rae69b}
R\"adler, K.-H. 1969b, Geod. Geophys. Ver\"offentlichungen Reihe II Heft 13, 131

\bibitem[R\"adler(1976)]{Rae76}
R\"adler, K.-H. 1976, In V. Bumba and J. Kleczek,
{\it Basic Mechanisms of Solar Activity},
D.\ Reidel Publishing Company Dordrecht, pp. 323--344

\bibitem[R\"adler et al.(2002a)]{Rae02a}
R\"adler, K.-H., Rheinhardt, M., Apstein, E., \& Fuchs, H., Magnetohydrodynamics 38, 39, 2002a

\bibitem[R\"adler et al.(2002b)]{Rae02b}
R\"adler, K.-H., Rheinhardt, M., Apstein, E., \& Fuchs, H., Nonlinear Processes in Geophysics 9, 171, 2002b

\bibitem[R\"adler et al.(2003)]{RKR03}
R\"adler, K.-H., Kleeorin, N., \& Rogachevskii, I.\ygafd{2003}{97}{249}

\bibitem[R\"adler et al.(2011)]{RBSR11}
R\"adler, K.-H., Brandenburg, A., Del Sordo, F., \& Rheinhardt, M.\ypr{2011}{E84}{046321}

\bibitem[Roberts \& Soward(1975)]{RS75}
Roberts, P. H., \& Soward, A. M.\yan{1975}{296}{49}

\bibitem[Rogachevskii et al.(2011)]{RKKB11}
Rogachevskii, I., Kleeorin, N., K\"apyl\"a, P. J., \& Brandenburg, A.\ypre{2011}{84}{056314}

\bibitem[R\"udiger \& Brandenburg(1995)]{RB95}
R\"udiger, G. \& Brandenburg, A.\yana{1995}{296}{557}

\bibitem[Schrinner et al.(2005)]{Sch05}
Schrinner, M., R\"adler, K.-H., Schmitt, D., Rheinhardt, M., Christensen, U.\yan{2005}{326}{245}

\bibitem[Schrinner et al.(2007)]{Sch07}
Schrinner, M., R\"adler, K.-H., Schmitt, D., Rheinhardt, M., Christensen, U. R.\ygafd{2007}{101}{81}

\bibitem[Steenbeck et al.(1966)]{SKR66}
Steenbeck, M., Krause, F., \& R\"adler, K.-H.\yjour{1966}{Z. Naturforsch.}{21a}{369}

\bibitem[Sur et al.(2008)]{SBS08}
Sur, S., Brandenburg, A., \& Subramanian, K.\ymn{2008}{385}{L15}

\bibitem[Vitense(1953)]{Vit53}
Vitense, E.\yjour{1953}{Z.\ Astrophys.}{32}{135}

\end{thebibliography}
\end{document}